%% file: 0_main.tex
\documentclass[conference]{IEEEtran}
\ifCLASSINFOpdf
\else
\fi

\usepackage{hyperref}

\usepackage{listings}
\makeatletter
\def\UrlAlphabet{%
      \do\a\do\b\do\c\do\d\do\e\do\f\do\g\do\h\do\i\do\j%
      \do\k\do\l\do\m\do\n\do\o\do\p\do\q\do\r\do\s\do\t%
      \do\u\do\v\do\w\do\x\do\y\do\z\do\A\do\B\do\C\do\D%
      \do\E\do\F\do\G\do\H\do\I\do\J\do\K\do\L\do\M\do\N%
      \do\O\do\P\do\Q\do\R\do\S\do\T\do\U\do\V\do\W\do\X%
      \do\Y\do\Z}
\def\UrlDigits{\do\1\do\2\do\3\do\4\do\5\do\6\do\7\do\8\do\9\do\0}
\g@addto@macro{\UrlBreaks}{\UrlOrds}
\g@addto@macro{\UrlBreaks}{\UrlAlphabet}
\g@addto@macro{\UrlBreaks}{\UrlDigits}
\makeatother
\usepackage{rotating}
\usepackage{xcolor}
\usepackage{longtable}
\usepackage{lscape}
\usepackage{tabularx}
\usepackage{lipsum}
\usepackage{booktabs}
\usepackage{colortbl}
\usepackage[normalem]{ulem}
\usepackage{ulem}
\useunder{\uline}{\ul}{}
\usepackage{multirow}
\usepackage{array}
\usepackage{utfsym}
\usepackage{bbding}
\usepackage{pifont}
\usepackage{makecell}
\usepackage{enumitem}
\usepackage{booktabs}
\usepackage{cleveref}
\usepackage{tablefootnote}
\usepackage[numbers]{natbib}
\usepackage{authblk}

\usepackage{diagbox}
\newcolumntype{Y}{>{\centering\arraybackslash}X} % 自动换行且水平居中
\newcolumntype{C}[1]{>{\centering\arraybackslash}m{#1}} % 固定宽度，内容垂直和水平居中

\usepackage{caption}
\usepackage{subcaption}
\definecolor{customblue}{HTML}{006ca6}
\definecolor{customgreen}{HTML}{009264}
\definecolor{custombrown}{HTML}{d4ac0d}
\AtEndPreamble{
 \usepackage{hyperref}
 \hypersetup{
  colorlinks = true,
  linkcolor = customblue,
  anchorcolor = customblue,
  citecolor = customgreen,
  filecolor = customblue,
  urlcolor = customblue
 }
}

\begin{document}

\title{A Solution toward Transparent and Practical AI Regulation: Privacy Nutrition Labels for Open-source Generative AI-based Applications}
% \author{\IEEEauthorblockN{Anonymous Authors}}

\author[1,*]{Meixue Si}
\author[2,3,*$^\dagger$]{Shidong Pan}
\thanks{1231213}
\author[3]{Dianshu Liao}
\author[3]{Xiaoyu Sun}
\author[2,3]{Zhen Tao}
\author[1]{Wenchang Shi}
\author[2,3]{Zhenchang Xing}
\affil[1]{\textit{Renmin University of China}}
\affil[2]{\textit{CSIRO's Data61}}
\affil[3]{\textit{Australian National University}}
\affil[*]{\small Equal contribution, $^\dagger$Corresponding author (shidong.pan@anu.edu.au)}

% \IEEEoverridecommandlockouts
% \makeatletter\def\@IEEEpubidpullup{6.5\baselineskip}\makeatother
% \IEEEpubid{\parbox{\columnwidth}{
%     Network and Distributed System Security (NDSS) Symposium 2024\\
%     26 February - 1 March 2024, San Diego, CA, USA\\
%     ISBN 1-891562-93-2\\
%     https://dx.doi.org/10.14722/ndss.2024.24xxx\\
%     www.ndss-symposium.org
% }
% \hspace{\columnsep}\makebox[\columnwidth]{}}

\maketitle

\begin{abstract}
    \input{0_abstract}
\end{abstract}

%\IEEEpeerreviewmaketitle

\input{1_Intro}
\input{2_StatusQuo}
\input{3_Motivation}

\input{4_PrivacyLabel}

\input{5_Repo2Label}

\input{6_Evaluation}
\input{7_Discussion}
\input{8_Conclusion}

\bibliographystyle{IEEEtranS}
\bibliography{ref}

\end{document}

%% file: 0_Abstract.tex
The rapid development and widespread adoption of Generative Artificial Intelligence-based (GAI) applications have greatly enriched our daily lives, benefiting people by enhancing creativity, personalizing experiences, improving accessibility, and fostering innovation and efficiency across various domains. 
However, along with the development of GAI applications, concerns have been raised about transparency in their privacy practices.
Traditional privacy policies often fail to effectively communicate essential privacy information due to their complexity and length, and open-source community developers often neglect privacy practices even more.
Only 12.2\% of examined open-source GAI apps provide a privacy policy.
To address this, we propose a regulation-driven GAI Privacy Label and introduce \texttt{Repo2Label}, a novel framework for automatically generating these labels based on code repositories.
Our user study indicates common endorsement of the proposed GAI privacy label format.
Additionally, \texttt{Repo2Label} achieves a precision of 0.81, recall of 0.88, and F1-score of 0.84 based on the benchmark dataset, significantly outperforming the developer self-declared privacy notices.
We also discuss the common regulatory (in)compliance of open-source GAI apps, comparison with other privacy notices, and broader impacts to different stakeholders.
Our findings suggest that \texttt{Repo2Label} could serve as a significant tool for bolstering the privacy transparency of GAI apps and make them more practical and responsible.

\vspace{5pt}

%% file: 1_Intro.tex
\section{Introduction}~\label{sec_intro}
%------------
%
%

\begin{figure*}[t]
    \centering
    \includegraphics[width=0.98\textwidth]{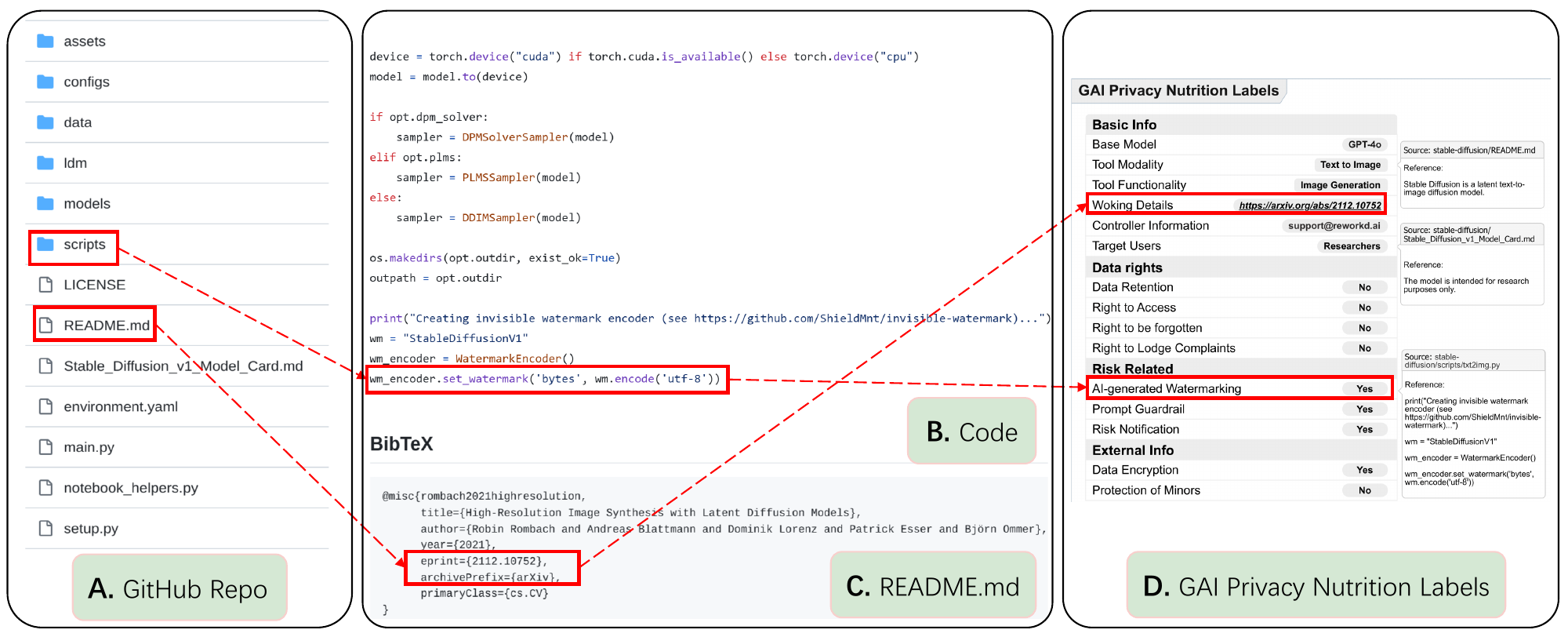}
    \caption{An overview of \texttt{Repo2Label} and an example GAI privacy label for \href{https://github.com/CompVis/stable-diffusion}{Stable Diffusion}. Given a repository (A), \texttt{Repo2Label} extracts all code files (B) and semi-structured textual documents (e.g., C) from the repository. Answers and references are then generated for each label filed in our proposed regulation-driven GAI privacy nutrition labels (D).}
    \label{fig:pnl_example}
\end{figure*}

%
%
%--------------
Online Generative Artificial Intelligence-based (GAI) applications\footnote{In the rest of the paper, we use ``GAI apps'' for short.} have emerged as powerful resources for a wide array of productive tools, ranging from content creation to decision support systems. 
These GAI apps often leverage powerful pre-trained large language models (LLMs) to generate new content that resembles the input data, enabling them to produce text, images, code, and more~\cite{fui2023generative}. 
The most typical tool is ChatGPT released in November 2022~\cite{openai2022}, which can create new content through the chatting interactions based on user prompts~\cite{mohamadi2023chatgpt,javaid2023study}. Since its release, ChatGPT attracted over one million users within just five days, and within two months, the user base surged to 100 million~\cite{milmo2023chatgpt}. 
Simultaneously, a plethora of open-source GAI apps have emerged, garnering significant attention within the open-source community. Empirical data indicate that the number of GAI repositories on the GitHub platform in 2023 has more than doubled compared to 2022~\cite{Octoverse}. These repositories often contain specific implementation code and guidance documentation for the GAI apps.
As an open-source and freely available project, AutoGPT has rapidly gained widespread attention within the open-source community.~\cite{firat2023if, ortiz2023auto, richards2023auto}. Remarkably, in just seven days, the project received 44,000 stars on GitHub~\cite{AutoGPTstars}. 

The rise of ChatGPT, AutoGPT, and other GAI apps that allow users to enter simple prompts has caused explosive growth among the public. 
The latest annual McKinsey Global Survey on the state of AI highlights the rapid expansion of GAI apps~\cite{McKinsey2023, McKinsey2024}. 79\% of all respondents report having been exposed to GAI, either for work or in other contexts. 
These GAI apps play increasingly crucial roles in both daily life and professional domains. 
However, considering the unique characteristics inherent to GAI, the complexity and diversity of their outputs may potentially give rise to trustworthiness concerns~\cite{sun2024trustllm,kenthapadi2023generative}, privacy concerns~\cite{golda2024privacy,carlini2021extracting,huang2022large, zhang2024forgotten, zhang2023right}, and copyright implications~\cite{carlini2023extracting, zhang2024privacy}. 
% \sd{you can also find some news (maybe from the New York Times) to support your statements}

Concurrently, governments worldwide have acknowledged the transformative impact of GAI and are proactively implementing measures to address the associated challenges. For instance, during the AI Seoul Summit\footnote{\url{https://www.gov.uk/government/topical-events/ai-seoul-summit-2024}} in May 2024, ten countries and the European Union reached an agreement to collaborate on the establishment of an international network dedicated to accelerating advancements in the science of AI safety~\cite{AISeoulSummit_10countries}, and 27 nations committed to work together on severe AI risks~\cite{AISeoulSummit_27nations}.
Policymakers have instituted a series of GAI-specific regulations aimed at safeguarding privacy, enhancing transparency, and ensuring the dependability of GAI apps. 
The recent amendments to Art. 52 of the EU AI Act~\cite{madiega2021artificial} have introduced specific regulations for GAI, mandating that AI systems generating deepfake content must disclose that the content has been artificially manipulated. Similarly, other countries such as Singapore~\cite{SingaporeGAI}, Canada~\cite{CanadaGAI}, and China~\cite{ChinaBasicRequirementsForGAI, ChinaMeasuresForGAI}  have also drafted and enacted legislation for GAI.
Additionally, it is important to recognize that GAI apps, as a specific subset of general software, should also comply with general regulations such as the General Data Protection Regulation (GDPR)~\cite{GDPR} and the California Consumer Privacy Act (CCPA)~\cite{CCPA}. 
These regulations, collectively seek to mitigate the risks associated with GAI deployment and foster its beneficial integration into society. 

To comply with transparency requirements mandated by privacy regulations, developers typically disclose software privacy practices to users through privacy notices. 
The most common form of these notices is the privacy policy~\cite{harkous2018polisis,cui2023poligraph,bui2023detection}. However, previous research~\cite{korunovska2020challenges,reidenberg2016ambiguity,singh2011evaluating} indicates that traditional privacy policies often suffer from excessive jargon, lengthy content, and ambiguous language, which undermine their effectiveness in securing informed consent from users~\cite{milne2004strategies}. To mitigate the problem of information overload in privacy policy communication, researchers have introduced a more concise, standardized, and easy-to-understand form of privacy notices -- \textit{privacy nutrition label}~\cite{kelley2009designing}. This \textit{privacy label} is designed to convey privacy information to users in a streamlined manner, enabling them to quickly and accurately access the details they are most concerned about.

In this study, we propose the regulation-driven \textit{GAI Privacy Nutrition Label}\footnote{In the rest of paper, we use "GAI Privacy Label" for short.} and a novel framework, dubbed \texttt{Repo2Label} (\textbf{\uline{Repo}}sitory s\textbf{\uline{to}} privacy nutrition \textbf{\uline{Label}}), to automatically generate GAI privacy labels based on the code repository (as shown in \autoref{fig:pnl_example}).
First, we conducted an empirical study of the status quo of GAI apps and their privacy notices. Only 12.2\% (18/148) of examined GAI apps offer a privacy policy, indicating a significant transparency deficiency in providing essential privacy information to end users. 
Then, we performed a thematic analysis of general privacy regulations (GDPR, CCPA, PIPL) and GAI-specific regulations, to establish a regulation-driven GAI privacy label format. 
Next, we introduced the design and implementation of \texttt{Repo2Label} framework which aims to automatically generate GAI privacy labels based on their code repositories.
In the evaluation, we evaluated various aspects of the proposed GAI privacy label design through a user study involving 48 participants. Results show that our proposed GAI privacy label format is widely endorsed by participants.
Additionally, based on the manual annotation dataset, \texttt{Repo2Label} achieves a precision of 0.81, recall of 0.88, and F1-score of 0.84 under the optimal experimental settings. 
Overall, the key contributions are:
\begin{itemize}
    \item To the best of our knowledge, this is the first research to empirically investigate the status quo of privacy notices to open-source GAI apps.
    \item To the best of our knowledge, this is the first research to propose regulation-driven privacy labels for GAI apps.
    \item We propose a \texttt{\textbf{Repo2Label}} framework for automatically generating GAI privacy labels based on code repositories. This code-based privacy notice generation method can more authentically reflect the privacy practices of GAI apps, compared to traditional self-declared approaches. 
\end{itemize}

The rest of this paper is organized as follows: In~\autoref{sec:2statusQuo}, we analyze the current state of GAI apps and their privacy notices.~\autoref{sec_motivation} presents the challenges faced by existing privacy labels and~\autoref{sec_label_design} introduces our regulation-driven GAI privacy nutrition label.~\autoref{sec:4method} details the \texttt{Repo2Label} framework for generating GAI privacy labels based on repositories.~\autoref{sec:5evaluation} reports a user study about our GAI privacy label design and describes the performance of the \texttt{Repo2Label} framework.~\autoref{sec:6discussion} discusses the broader impact of our work on the community, and we conclude in~\autoref{sec:7conclusion}.

Ethical approval for this research was secured
from our institution’s Institutional Review Board (IRB).

%% file: 2_StatusQuo.tex
\section{Status Quo of GAI Applications}~\label{sec:2statusQuo}
This section describes our empirical observations of GAI apps and an in-depth analysis of the current GAI apps on the market. Additionally, the problems existing in the current privacy notices, especially privacy policies, of GAI apps are discussed.

\subsection{GAI-based Applications}
% definition要对应label里面的内容：modality、functionality
Generative Artificial Intelligence~\cite{weisz2023toward} refers to AI systems that utilize existing media to generate new content~\cite{nigam2021beyond,saleem2020generative,sbai2018design}. These systems leverage large datasets to learn patterns and structures, enabling them to create original outputs that resemble the input data. This capability spans various modalities, including text, images, audio, and video, making GAI a versatile tool in numerous apps~\cite{lyu2023design}. A variety of apps with different functionality are popping up on the market. For instance, it can be used to produce realistic images from textual descriptions, generate emails, write news, and even simulate voices. The potential of GAI lies in its ability to extend human creativity and productivity by automating the creation of high-quality, innovative content.

Recent advancements in GAI, especially in the domain of LLMs, have substantially improved the ability of these systems to understand and process textual information. The leaps in this field have empowered GAI apps to interpret user inputs with greater precision and generate contextually relevant responses. This progress has markedly reduced the accessibility barriers to such GAI apps, enabling users, even those with limited expertise in prompts or instructions, to effectively utilize text-guided generation capabilities~\cite{ali2024constructing}. 

% In this study, we define the GAI applications by the following rules: 1) driven by a powerful foundation GAI model; 2) an online interaction interface without local deployment; and 3) no compulsory monetary spending is required to use. This is the most common set of GAI applications that are available to normal users.

% Recent surveys have shown that there is widespread use of GAI and a growing demand for these applications.

% GAI denotes a category of AI systems capable of creating new content through text, images, or other forms of media. GAI represents a dramatic advancement from previous AI models. GAI leverages deep learning models to generate human-like content, including audio, code, images, text, simulations, 3D objects, and videos. These applications can create unexpected outputs in response to varied and complex prompts (e.g., languages, instructions, questions). 
% GAI has already shown itself to be incredibly useful for planning vacations, answering questions, researching, getting product information, learning about medical conditions, understanding data, and generating images, videos, and computer code. It can even write customized on-demand poetry and song lyrics. 

\subsection{An Overview of GAI Application Market} \label{sec_market_app_analysis}
% According to GitHub's official 2023 annual Octoverse report~\cite{Octoverse}, there were 65,000 new GAI projects added in 2023, representing a 248\% year-over-year increase~\cite{dohmke2023sea}. This surge contributed to an overall annual growth of 27\% in the total number of projects on GitHub. These figures indicate the rapid proliferation of GAI applications within the open-source community.

% As public interest in GAI applications escalates, these applications are increasingly classified with a fine degree of granularity according to their intended applications, such as Email Generation, Advertisement Generation, and Recipe Generation, among others. Each category of application is imbued with substantial capabilities, tailored to perform specific tasks with high efficiency.

\input{table/2.2Market_Analysis_of_GAI_Tools}

Third-party GAI app collection websites are independent online marketplaces established by external developers or organizations.
The gpt3demo\footnote{\href{https://gpt3demo.com}{https://gpt3demo.com}} and gpt4demo\footnote{\href{https://gpt4demo.com}{https://gpt4demo.com}} are two popular third-party websites that actively collect and curate GAI apps and demonstrations since the mid of 2022, as pioneers of their kind.
We use those two popular third-party collections to harvest mainstream GAI apps on the market.
Through the deployment of customized web scrapers, we successfully retrieved information including the name, category tags, code repository link, and privacy notices (e.g., privacy policies).
For the gpt3demo site, a total of 887 GAI apps are categorized into 228 fine-grained categories. Meanwhile, the gpt4demo site lists 87 GAI apps, spread across 47 distinct categories. 
Most of these GAI apps position themselves as domain experts, incorporating professional domain knowledge to act as advisors for users in specific fields, including but not limited to Robot Lawyers and Coding Assistants. 
% Among these GAI applications, the categories of \textit{Developer Applications}, \textit{GPT-3 Alternative LLMs}, and \textit{AI Copywriting} are the top three most popular, with totals of 81, 63, and 54 applications respectively (as illustrated in Figure~\ref{fig:numbers of GAI applications}). 
We manually checked and selected open-source GAI apps that explicitly provided a GitHub repository address. 
After excluding three inaccessible repositories, we successfully collected 148 open-source GAI apps with a valid GitHub repository link, with the proportions being 15.2\% of all curated GAI apps.
As shown in~\autoref{table_pp_statistics}.c, the average stars and the average forks are 11.4k and 1.9k, respectively, reflecting the significant popularity of those open-source GAI Apps.
According to Repositories Ranking~\cite{reporank}, the median stars of the top 10,000 GitHub repositories is about 5.8k.
GAI apps provide expertise and consultation to support a wide range of app scenarios. 
However, the complexity and capability of these apps lead to potentially significant privacy risks, which remain to be adequately addressed~\cite{zhang2024privacy}.

% \begin{figure}[h]
%     \centering
%     \includegraphics[width=0.50\textwidth]{figure/numbers_of_GAI_tools.pdf}
%     \caption{Tags of GAI tools in gpt3demo and gpt4demo. Tags with less than seven tools are omitted.}
%     \label{fig:numbers_of_GAI_tools}
% \end{figure}

Notably, as this is a super dynamic market, our data collection is a snapshot from January 2024. This subset is sufficient to demonstrate some characteristics of GAI apps discussed in this section.
OpenAI official GPTs store\footnote{\url{https://chatgpt.com/gpts}} is another GAI app repository that attracts increasing attention, but all apps are solely based on GPT foundation models and most of them are not open-source.

%--------------
%
%
\begin{figure}[t]
    \centering
    \includegraphics[width=1\linewidth]{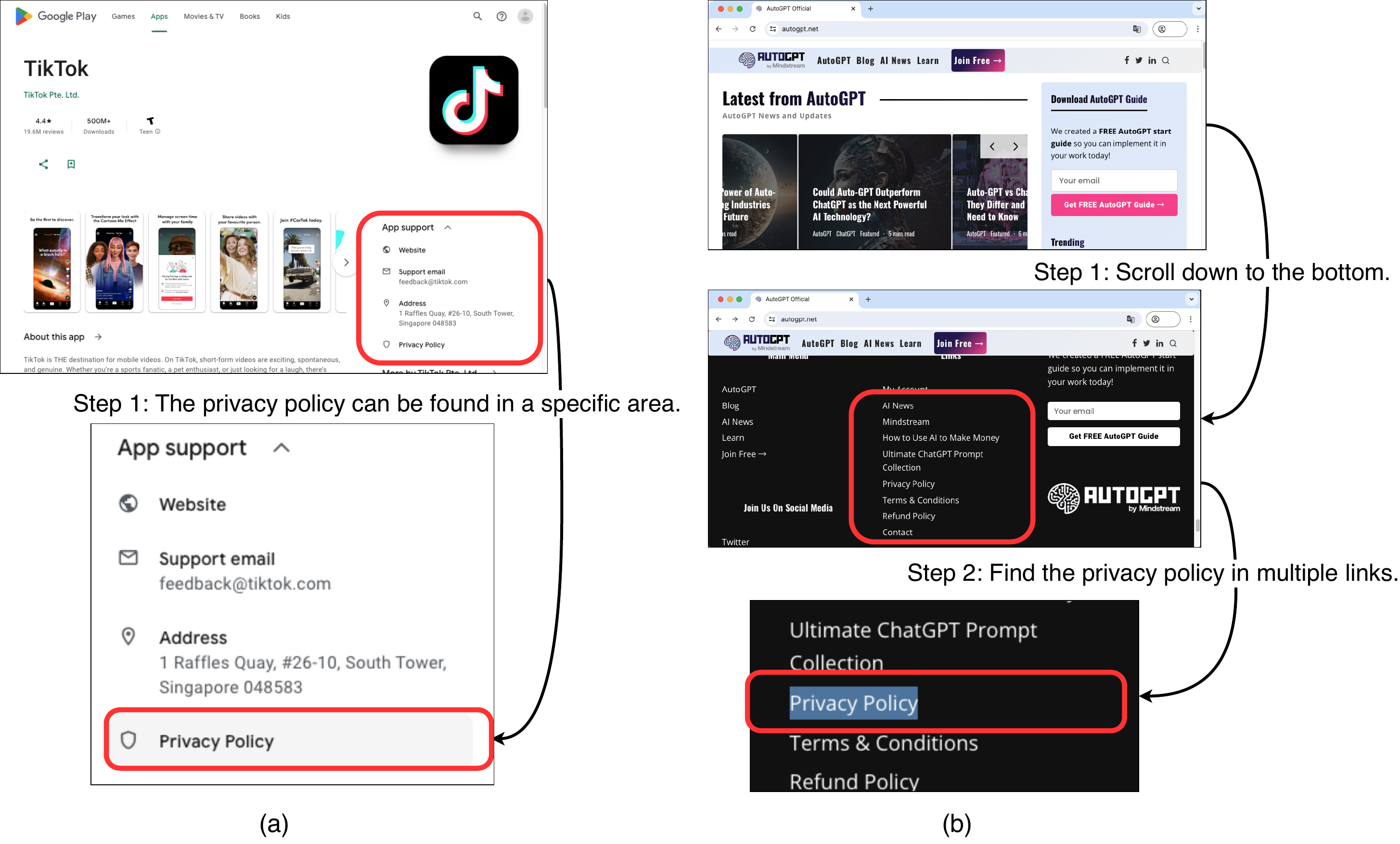}
    \caption{(a) The position of privacy policies of mobile apps in the Google Play app store. (b) The position of the privacy policy of AutoGPT, one of the GAI apps.}
    \label{fig:PP_comparison}
\end{figure}
%
%
%--------------

\subsection{Current Privacy Notices of Open-source GAI Applications}
The privacy notice is an essential component of software, including the emerging GAI apps and tools. 
The privacy policy is the most common type of privacy notice and is widely required by privacy regulations and industry standards~\cite{pantrap, pan2023seeprivacy}.
In the mobile app ecosystem, two of the largest app stores, Google Play and Apple App Store, both mandate that app developers provide a privacy policy before publishing an app.
\autoref{fig:PP_comparison}.a shows that the privacy policies of mobile apps in the Google Play app store are curated in a visually obvious and easily accessible position (in the sidebar) on the homepage.
However, the GAI apps market is an emerging sector, and there are no such industry standards to enforce developers to follow. Therefore, we conduct an empirical study to investigate the current state and practices of privacy notices for open-source GAI apps, as follows.

For the 148 GAI apps, we conducted manual scrutinization about their GitHub repositories, official websites (if available), and tool interfaces (if available) to check the provision of privacy notices, especially privacy policies.
Specifically, we visited all the aforementioned websites and searched for relevant information in their HTML sources. 
The keywords used for the search included \textit{``privacy'', ``privacy policy'', ``notices'', ``terms'', ``terms of services''}, etc. Although the terms of services (ToS) are not privacy notices, they are commonly put aside in practice. Those keywords might help us to find privacy notices.
Additionally, we employed the Google search engine to broadly search the privacy notices for GAI apps.
The search terms are ``[GAI application name] + [keyword] (e.g., privacy policy)'', and for each search keyword, we evaluated all entries on the first result page.
For example, the privacy policies of AgentGPT\footnote{Its official website is \url{https://agentgpt.reworkd.ai}, and its privacy policy is available at \url{https://agentgpt.reworkd.ai/privacypolicy.html}}
and PromptLayer\footnote{Its official website is \url{https://promptlayer.com/}, and its privacy policy is available at \url{https://promptlayer.com/privacy_policy.pdf}}
are discovered by additional searching, but their privacy policies are not directly included in their repositories or main interfaces. \autoref{fig:PP_comparison}.b shows the privacy policy link of AutoGPT, located in the footer of the advertisement website, is not easily accessible to users.
In total, only 12.2\% (18/148) of examined GAI apps offer a privacy policy as listed in \autoref{table_pp_statistics}.a, indicating a significant transparency deficiency in providing essential privacy information to end users. 
Our findings also highlight the issue that existing privacy policies can be difficult for users to find without extra effort.

Among the small portion of GAI apps, the readability of their privacy policies is as unsatisfactory as reported in previous studies~\cite{pantrap, li2022understanding}. These policies are often lengthy, filled with jargon, and contain frequent hyperlinks and cross-references.
\autoref{table_pp_statistics}.b shows the statistics of GAI app privacy policies we collected.
The average length of examined privacy policies is about 3,000 words, and the average reading time is about 12 minutes, according to~\cite{Blakkarly2022privacy}.
We then calculated the readability based on the Flesch Reading-Ease Test for privacy policies~\cite{kincaid1975derivation, pantrap}, and the average readability score is 39.9.
This number indicates that fully comprehending the privacy policy of GAI apps requires at least a college-level education~\cite{kincaid1975derivation}, which contradicts the mission of GAI apps to mitigate the expertise gap.

%If internet users were to actually read the privacy policies of the websites they visit, it would take them approximately 250 hours per year ~\cite{mcdonald2008cost}.
%The most significant flaws were their lack of readability and transparency~\cite{milne2004strategies}

Surprisingly, through additional searching, we noticed only two (1.3\%) GAI apps provide privacy labels as their privacy notices, instead of privacy policies. 
Privacy labels are short-form, clear, table-like disclosures that enable users to quickly understand how their data is collected and utilized~\cite{ kelley2009nutrition, kelley2013privacy, emami2020ask}.
Privacy labels have been proven useful over the past decade, and the concept has now been widely applied in the industry, such as Apple and Google.
Comparably, 60\% of apps on the Apple App Store and 44\% of apps on the Google Play Store have completed the necessary forms to generate privacy labels, as of August 26, 2022~\cite{li2024matcha}.
The two existing privacy labels for GAI apps are compulsorily required privacy disclosures (similar to the Data Safety section in the Google Play app store) in the Chrome web store\footnote{\url{https://chromewebstore.google.com/category/extensions}} for Chrome extensions, as both GAI apps are implemented and launched in the Chrome store.
Such privacy label format is tailored for browser extensions, neglecting the unique challenges and regulatory requirements of GAI apps. 
We detail the current challenges of privacy labels in the next section and propose a GAI-specific privacy label format in \autoref{sec_label_design}.

%Many researchers have been investigating ways to make privacy notices more accessible for users. Proposals such as comic-based privacy policies~\cite{tabassum2018increasing}, privacy icons~\cite{habib2021toggles, efroni2019privacy, cranor2020effectively}, and privacy labels~\cite{kelley2009nutrition} have been put forth. The most profound impact has been made by privacy labels. 
%Companies such as Apple, Google, and Amazon require their developers to use privacy labels. This privacy notice appears on the public page of every application and is self-reported by the application developers. As of August 26, 2022, 60\% of applications on the Apple App Store and 44\% of applications on the Google Play Store have completed the necessary forms to generate these labels~\cite{li2024matcha}. This privacy label has also been utilized in the realm of Internet of Things (IoT) devices~\cite{emami2020ask}.
% Privacy labels hold immense potential in maintaining software transparency. Yet, based on our collected dataset of GAI applications, only two of the GAI applications offer privacy labels (Table \ref{table:demo.com statistics}\dianshu{Please revise this type Table ??}), and both are extensions in the Chrome app store. The coverage of privacy labels is \sd{X\%.}

%% file: table/2.2Market_Analysis_of_GAI_Tools.tex
\begin{table}
\scriptsize
\caption{The statistic analysis of GAI Apps on gpt3demo.com and gpt4demo.com.}
\label{table_pp_statistics}
\begin{minipage}{\textwidth}
\begin{subtable}{0.5\textwidth}
\centering
\begin{tabular}{lccc}
\toprule
\textbf{} & \textbf{gpt3demo} & \textbf{gpt4demo} & \textbf{Total} \\ \midrule
Available Code Repository & 138 & 10 & 148 \\
%Relevant Website & 60 & 7 & 67 \\
Available Privacy Policy & 16 & 2 & 18 \\ 
Available Privacy Labels & 1 & 1 & 2 \\ 
\midrule
GAI Tools & 887 & 87 & 974 \\ 
\bottomrule
\end{tabular}
\subcaption{Overall.}
\end{subtable}
\end{minipage}
\begin{minipage}{\textwidth}
\begin{subtable}[t]{0.25\textwidth}
\centering
\scriptsize
\label{table:PP statistics}
\begin{tabular}{lr}
\toprule
\textbf{No. Privacy policies} & \textbf{18} \\ \midrule
No. Words & 53,540 \\
No. Sentences & 2,222 \\
Avg. Words per Sentence & 2,974 \\
Avg. Sentences per Privacy policy& 123 \\ \bottomrule
\end{tabular}
\subcaption{Privacy policies.}
\end{subtable}
\hspace{0.01\textwidth}
\begin{subtable}[t]{0.25\textwidth}
\centering
\label{table:stars statistics}
\begin{tabular}{lr}
\toprule
\textbf{No. Repositories} & \textbf{148} \\ \midrule
No. Stars & 1,693.7k \\
No. Forks & 288.3k \\
Avg. Stars & 11.4k \\
Avg. Forks & 1.9k \\ \bottomrule
\end{tabular}
\subcaption{GitHub Repositories.}
\end{subtable}
\end{minipage}
\end{table}

%% file: 3_Motivation.tex
%-----------
%
%
\input{table/0701regulation-empirical_study}
%
%
%-----------

\section{Motivation}~\label{sec_motivation}
In this section, we examine the challenges encountered in creating these privacy labels.
Privacy labels are designed to accurately convey the software privacy practices to users in a timely manner. 
In the realm of software ecosystems, the implementation of extensive privacy labels ostensibly enhances transparency and user comprehension regarding data handling practices. However, researchers and consumer advocates have articulated various challenges regarding the present privacy labels and the generation~\cite{xiao2023lalaine, li2022understanding, zhang2022usable, kollnig2022goodbye, li2022understanding1, koch2022keeping}.

% 对于第二点，~\cite{lin2023data}的工作比较了ios和安卓隐私标签的不同。ios分为四个部分，安卓分为两类。隐私标签的可用性存在问题（how usable are ios app privacy labels?）~\cite{zhang2022usable}
\textbf{Challenge-1.} The concept of privacy labels has transitioned from theoretical frameworks to practical apps. 
Apple~\cite{campbell2020apple}, Google~\cite{porter2022google}, and Amazon~\cite{AmazonPL} have sequentially mandated developers to provide privacy labels for apps hosted within their respective app marketplaces. 
%Apple pioneered this initiative in December 2020, stipulating the inclusion of privacy labels for iOS applications. Subsequently, Google introduced mandatory privacy labels, termed as ``Data Safety'' labels, for Android applications in July 2022. In April 2023, Amazon also joined this movement, necessitating the provision of privacy labels for applications, and displaying a notification on application detail pages for those without such labels, stating ``Information not provided by developer''.
% Privacy labels are prominently featured on application download pages, aimed at enhancing user awareness regarding the types of data collected by an application. This not only augments application transparency but also facilitates informed decision-making among users when selecting and utilizing applications. 
However, despite the widespread adoption of privacy labels at the corporate level, a lack of uniformity and established standards persists throughout their implementation processes. Previous work has discussed the differences in privacy labels between Apple and Google. Lin et al.~\cite{lin2023data} conducted a detailed comparison of the privacy label designs of Apple and Google. Their study identified significant differences in the structure and categorization of data within the privacy labels of the two companies. Additionally, the research highlighted that the two companies use different terminologies to describe the same concepts.
Cranor et al.~\cite{cranor2022mobile} have discussed the missing key ingredients for mobile app privacy labels. 
The lack of standardization and consistency in privacy label design remains a significant obstacle. 
This inconsistency not only hampers users' ability to make informed decisions regarding their data privacy but also complicates regulatory compliance for developers.

\textbf{Challenge-2.} 
A central concern pertains to the accuracy of the labels, questioning their fidelity to accurately represent privacy practices. 
The primary rationale behind this issue stems from the fact that the privacy label generation mainly relies on questionnaire-based methodologies~\cite{li2022understanding, pan2023toward, li2024matcha}. 
This process involves querying app developers with a set of inquiries regarding the privacy aspects of their apps and subsequently utilizing their responses to generate privacy labels.
Self-declaration can create privacy labels, but their quality could vary~\cite{li2022understanding, pan2023toward, li2024matcha}. 

Developing privacy labels is a complex process, demanding both the knowledge of app features and corresponding legal requirements. 
From the perspective of developers, the task of meticulously completing the requisite questionnaires for the generation of a privacy label presented a considerable challenge.
First, developers often hold misconceptions regarding the terminology of privacy labels~\cite{li2022understanding}; In addition, they might not completely understand the behavior of software, especially parts from collaborators in the development team.
Therefore, it is important to propose a code-based privacy label generation approach to accurately reflect the actual behaviors of GAI Apps.
%In the design and implementation of software systems, privacy considerations are frequently relegated to a secondary position\cite{li2021developers, balebako2014improving}. This prioritization is primarily due to the substantial resources required for effective privacy protection, encompassing time, manpower, and financial investment. The significant additional costs associated with these resources often result in privacy protection being perceived as an extraneous burden rather than an integral component of the core development tasks.

% \mx{so many third-party libraries, memory limitations, code modification}
% Given the rapid iteration of GAI tools, developers often struggle to retain awareness of privacy-related changes precipitated by code modifications. The development of GAI tools involves multiple collaborators; consequently, the creation of more precise privacy labels necessitates coordination among several colleagues, which is highly challenging. In particular, AI software often relies on third-party libraries and APIs to access data, perform computations, and enhance functionality. However, due to the limitations of developers' memory capacity, thoroughly understanding and retaining knowledge of the data practices of third-party libraries presents a significant challenge.

% Privacy labels allowed users to better compare privacy practices between different applications, and the studies found standardized formats more enjoyable to read. Standardized labels could increase both the speed of finding information and the accuracy of users’ comprehension.

%% file: table/0701regulation-empirical_study.tex
\begin{table*}[ht]
\caption{Thematic analysis results for requirements in general privacy regulations and GAI-specific regulations. Meas-GAI denotes Administrative Measures for Generative Artificial Intelligence Services~\cite{ChinaMeasuresForGAI}. Req-GAI denotes Basic Security Requirements for Generative Artificial Intelligence Service~\cite{ChinaBasicRequirementsForGAI}. Prin-GAI denotes Principles for Responsible, Trustworthy and Privacy-Protective Generative AI Technologies ~\cite{CanadaGAI}. MAIF-GAI denotes Model AI Governance Framework for Generative AI~\cite{SingaporeGAI}. In the subsequent design of GAI privacy label, we further elaborate \textit{Tool Type} into \textit{Tool Modality} and \textit{Tool Functionality}. Notably, we only include the intersection of these requirements in this table.}
% \footnotesize
\centering
\begin{tabular}{llcll}
\toprule
\textbf{Regulation} & \textbf{Region} & \multicolumn{1}{c|}{\textbf{\begin{tabular}[c]{@{}c@{}}Publish\\ Date\end{tabular}}} & \multicolumn{2}{c}{\textbf{Operationalized Requirements [Article Reference]}} \\ \midrule
\multicolumn{5}{c}{\cellcolor[HTML]{C0C0C0}\textit{General Privacy Regulations}} \\ \midrule
GDPR\cite{GDPR} & EU & \multicolumn{1}{c|}{Apr'16} & Right to Lodge Complaints [Art.13.2.(d) \& 14.2.(e)] & Data Encryption [Art.32.1.(a)] \\
 &  & \multicolumn{1}{c|}{} & Right to be Forgotten [Art.13.2.(c) \& 14.2.(d)] & Data Retention [Art.13.2.(a) \& Art.14.2.(a)] \\
 &  & \multicolumn{1}{c|}{} & Right to Access [Art.13.2.(b) \& 14.2.(c)] & Right to Lodge Complaints [Art.13.2.(e)] \\
 &  & \multicolumn{1}{c|}{} & Controller Contact [Art.13.1.(a)] & Protection of Minors[Art.32.1] \\ \midrule
CCPA\cite{CCPA} & California & \multicolumn{1}{c|}{Jun'18} & Right to be Forgotten [\textsection{}1798.120] & Right to Access [\textsection{}1798.110] \\
 &  & \multicolumn{1}{c|}{} & Data Retention [\textsection{}1798.100.a.(3)] &  \\ \midrule
PIPL\cite{PIPL} & China & \multicolumn{1}{c|}{Aug'21} & Right to be Forgotten [Art.15] & Data Encryption [Art.51.(3)] \\
 &  & \multicolumn{1}{c|}{} & Controller Contact [Art.17.(1) \& Art.52] & Right to Access [Art.45] \\
 &  & \multicolumn{1}{c|}{} & Data Retention [Art.17.(2) \& Art.19] & Risk Notification [Art.51] \\
 &  & \multicolumn{1}{c|}{} & Right to Lodge Complaints [Art.50] & Protection of Minors [Art.31] \\ \midrule
\multicolumn{5}{c}{\cellcolor[HTML]{C0C0C0}\textit{GAI-specific Regulations}} \\ \midrule
Meas-GAI\cite{ChinaMeasuresForGAI} & China & \multicolumn{1}{c|}{May'23} & Base Model [Art.7] & Target Users [Art.10] \\
 &  & \multicolumn{1}{c|}{} & Protection of Minors [Art.10] & Tool Type [Art.10] \\
 &  & \multicolumn{1}{c|}{} & AI-generated Watermarking [Art.12] & Data Retention [Art.11] \\
 &  & \multicolumn{1}{c|}{} & Right to Lodge Complaints [Art.15 \& Art.18] & Risk Notification [Art.14] \\ \midrule
Req-GAI\cite{ChinaBasicRequirementsForGAI} & China & \multicolumn{1}{c|}{Oct'23} & Prompt Guardrail [Art.6.(b).1 \& Art.7.(f).1 \& Art.6.(b).2] & AI-generated Watermarking [Art.7.(d)] \\
 &  & \multicolumn{1}{c|}{} & Right to Lodge Complaints [Art.5.2.(b).3 \& Art.7.(e)] & Tool Type [Art.6.(c).1] \\
 &  & \multicolumn{1}{c|}{} & Risk Notification [Art.5.2.(b).4 \& Art.6.(b).2] & Target Users [Art.6.(c).1] \\
 &  & \multicolumn{1}{c|}{} & Base Model [Art.6.(a) \& Art.6.(c).1 \& Art.6.(c).2] & Right to be Forgotten [Art.7.(c)] \\
 &  & \multicolumn{1}{c|}{} & Protection of Minors [Art.7.(a).3 \& Art.7.(a).4] &  \\ \midrule
Prin-GAI\cite{CanadaGAI} & Canada & \multicolumn{1}{c|}{Dec'23} & Data Encryption [Art.3] & Tool Type [Art.4] \\
 &  & \multicolumn{1}{c|}{} & AI-generated Watermarking [Art.4] & Right to Access [Art.6] \\
 &  & \multicolumn{1}{c|}{} & Risk Notification [Art.4] & Data Retention [Art.7] \\
 &  & \multicolumn{1}{c|}{} & Working Details [Art.5] & Prompt Guardrail [Art.8] \\ \midrule
MAIF-GAI\cite{SingaporeGAI} & Singapore & \multicolumn{1}{c|}{Jan'24} & Prompt Guardrail [Art.3.(d) \& Art.3] & Base Model [Art.3.(b)] \\
 &  & \multicolumn{1}{c|}{} & Risk Notification [Art.3.(e) \& Art.6.(a)] & Tool Type [Art.3.(f)] \\
 &  & \multicolumn{1}{c|}{} & AI-generated Watermarking [Art.7] & Working Details [Art.3.(d) \& Art.3.(g)] \\ \midrule
AI Act\cite{madiega2021artificial} & EU & \multicolumn{1}{c|}{TBD} & AI-generated Watermarking [Art. 52] &  \\ \midrule
\end{tabular}

\label{table:regulations empirical study}
\end{table*}

%% file: 4_PrivacyLabel.tex
\section{Proposed GAI Privacy Labels}\label{sec_label_design}
To respond the Challenge-1 discussed in~\autoref{sec_motivation}, we conduct a \textbf{regulations-driven} privacy label design process and propose a GAI privacy label format. We then evaluate the proposed design through a human evaluation in \autoref{sec_eva_label_design}.

% \subsection{GAI Privacy Label Design}

In the design and implementation of privacy labels, it is crucial to identify the key components that should be included. 
The concept of privacy labels was first inspired by the nutritional labels on food packaging and introduced by Kelley et al.~\cite{kelley2009nutrition} in 2009. 
After that, various types of privacy labels have been proposed for different scenarios, including websites~\cite{kelley2010standardizing}, IoT devices~\cite{emami2020ask}, and mobile apps~\cite{pan2023toward, porter2022google, campbell2020apple}.
Although the format and information in those privacy labels are different, they are commonly constituted by three major sections: a) data controller information; b) data practices and purposes; and c) risk disclosures.

%After more than a decade of development, this concept has evolved from a theoretical framework to practical application in the market. Recently, major technology companies such as Apple~\cite{campbell2020apple}, Google~\cite{porter2022google}, and Amazon have mandated that developers provide privacy labels for applications available in their app stores. However, the current implementation of these privacy labels varies across different platforms, with each company adopting its own standards for format and content~\cite{lin2023data}. Developers typically generate these labels through self-reported forms, a process that can lead to inaccuracies due to misunderstandings of terminology or intentional misreporting~\cite{li2022understanding,li2022understanding1,xiao2023lalaine}. The lack of a standardized format further complicates users' ability to understand and compare privacy information across different applications~\cite{lin2023data}.

To tackle the aforementioned issues, we aim for a standardized GAI privacy label to not only transparently, but also compliantly disclose the privacy practices of GAI apps. 
To this end, we conducted an empirical study of existing regulations to examine and draft a GAI privacy label format.
GDPR~\cite{GDPR} and CCPA~\cite{CCPA} are pioneers on the general data and privacy protection, followed by them, China launched the PIPL~\cite{PIPL} to fulfill the blank.
With the development of Generative AI models, GAI-based apps present unique challenges not seen in previous scenarios. 
First, it is difficult to finely define data types in GAI app interactions, as most GAI apps take diverse and sophisticated prompts from users as inputs. Unlike traditional scenarios where data types are relatively straightforward, interactions with GAI apps can involve nuanced and context-dependent data.
Second, GAI apps often process large volumes of personal and sensitive data in different modalities, and their advanced capabilities can infer additional information about users. 
This introduces heightened privacy risks, as the potential for data misuse or unintended consequences increases. 
Third, data rights, such as the Right to Access, Rectify, or be Forgotten, become more complex in the context of GAI, and have attracted increasing attention from users.

To respond to the unique challenges, governments have placed GAI-specific regulation legislation on the agenda.
By October 2023, 31 countries had passed AI legislation, and 13 more were debating AI laws~\cite{AIRegCount}.
These GAI-specific regulations impose comprehensive requirements on GAI apps, focusing on risk assessment and disclosure to ensure compliance with legal standards.
In total, we consider three general privacy regulations (GDPR, CCPA, PIPL) and five GAI-specific regulations in four countries/regions (China, Canada, Singapore, and the EU).
For each regulation, we carefully scrutinized and extracted requirements about GAI apps. 
For instance, Singapore \textit{Model AI Governance Framework for Generative AI}~\cite{SingaporeGAI}, article 3, stipulates that ``\textit{A crucial step for safety is also to consider the context of the use case and conduct a risk assessment. For example, further fine-tuning or using user interaction techniques (such as input and output filters) can help to reduce harmful output...}''. 
Upon conducting a thematic analysis of this requirement, we have encapsulated it as \textit{Prompt Guardrail}, which is then incorporated as a GAI privacy label field. 
Two authors conducted the thematic analysis and created the initial codebook, independently. 
Both authors have at least two years experience on privacy regulation, AI governance, and Responsible AI field. 
For any disagreement in the codebook, they discussed and agreed on the same answer, and if the disagreement persisted, a third author joined the discussion to facilitate a resolution.
Above all, we determined the regulation-driven GAI privacy label fields as shown in \autoref{table:regulations empirical study}.  
%-------
%
%
\input{table/Tallies}
%
%
%-------
\autoref{table:Tallies} presents the existence for each label field across various regulations.
In the design of GAI privacy label format, we meticulously considered the limitations of human cognitive psychology, particularly the concept of information chunking as proposed by Miller~\cite{miller1956human,miller1956magical}.
\autoref{fig:pnl_example}.D illustrates an example of our proposed GAI privacy label. To optimize user comprehension and retention, we categorized the privacy label items into four primary sections:
\begin{enumerate}
    \item \textbf{Basic Info}: This section aims to provide a fundamental description of the GAI app, including its base model, supported modalities, primary functions, working details, developer information, and target users.
    \item \textbf{Data Rights}: This section details the essential data rights users possess when utilizing the tool, such as the right to access and the right to be forgotten.
    \item \textbf{Risk Related}: This section outlines any potential risks associated with the use of the tool, including whether there are risk notifications and if there are identifiers for content generated by the tool.
    \item \textbf{Additional Info}: This section discloses information on data encryption and special protections for minors.
\end{enumerate}
The explanations for each label item are shown in \autoref{table:label explanation}.
%\sd{For the \textit{Tool Modality}, \textit{Tool Functionality}, and \textit{Target User} in Basic Info, we determined all potential answers using a snowballing method. To make the answer list snowball, we manually checked the GPT3demo and GPT4demo GAI apps. We stopped when no new keywords appeared in 50 consecutive GAI apps.}
Additionally, the last three groups of privacy label fields have binary content. ``Yes'' denotes that this GAI app does implement this requirement, and ``No'' denotes otherwise.
For each label item, there will be a floating bubble explaining the source of each answer.
This structured approach ensures that users are comprehensively informed about the privacy aspects of the tool, enhancing transparency and trust.

%Above all, to address the absence of a uniform standard for GAI privacy labels, we conducted a regulation-driven empirical study and proposed our design.
%Based on our findings, we drafted a new Generative Artificial Intelligence (GAI) nutrition label format. This proposed privacy label format aims to enhance the transparency of generative AI tools, providing a valuable reference for further research in this area.

\input{table/Explanations}

% \begin{figure}[h]
%     \centering
%     \includegraphics[width=0.50\textwidth]{figure/Privacy_Nutrition_Labels.pdf}
%     \caption{The format of proposed GAI privacy label and an example of \href{https://github.com/CompVis/stable-diffusion}{Stable Diffusion}.}
%     \label{fig:label design}
% \end{figure}

%% file: table/Tallies.tex
\begin{table*}
\caption{Tallies of GAI privacy label fields according to regulations.}
\centering
\rowcolors{2}{gray!10}{white}
\resizebox{\textwidth}{!}{
\begin{tabular}{l|ccc|ccccc|c}
\hline
\textbf{} & \multicolumn{3}{c|}{\textbf{General Regulations}} & \multicolumn{5}{c|}{\textbf{GAI-specific Regulations}} &  \\ \hline
\diagbox[width=12em]{Label Field}{Regulations}{Country} & \begin{tabular}[c]{@{}c@{}}EU\\ GDPR\end{tabular} & \begin{tabular}[c]{@{}c@{}}California\\ CCPA\end{tabular} & \begin{tabular}[c]{@{}c@{}}China\\ PIPL\end{tabular} & \begin{tabular}[c]{@{}c@{}}China\\ Meas-GAI\cite{ChinaMeasuresForGAI}\end{tabular} & \begin{tabular}[c]{@{}c@{}}China\\ Req-GAI\cite{ChinaBasicRequirementsForGAI}\end{tabular} & \begin{tabular}[c]{@{}c@{}}Canada\\ Prin-GAI\cite{CanadaGAI}\end{tabular} & \begin{tabular}[c]{@{}c@{}}Singapore\\ MAIF-GAI\cite{SingaporeGAI}\end{tabular} & \begin{tabular}[c]{@{}c@{}}EU\\ AI Act\cite{madiega2021artificial}\end{tabular} & Tallies \\ \hline
Base Model & \ding{55} & \ding{55} & \ding{55} & \Checkmark & \Checkmark & \ding{55} & \Checkmark & \ding{55} & 3/9 \\
Tool Type & \ding{55} & \ding{55} & \ding{55} & \Checkmark & \Checkmark & \Checkmark & \Checkmark & \ding{55} & 4/9 \\
Working Details & \ding{55} & \ding{55} & \ding{55} & \ding{55} & \ding{55} & \Checkmark & \Checkmark & \ding{55} & 2/9 \\
Controller Contact & \Checkmark & \ding{55} & \Checkmark & \ding{55} & \ding{55} & \ding{55} & \ding{55} & \ding{55} & 2/9 \\
Target Users & \ding{55} & \ding{55} & \ding{55} & \Checkmark & \Checkmark & \ding{55} & \ding{55} & \ding{55} & 2/9 \\
Data Retention & \Checkmark & \Checkmark & \Checkmark & \Checkmark & \ding{55} & \Checkmark & \ding{55} & \ding{55} & 5/9 \\
Right to Access & \Checkmark & \Checkmark & \Checkmark & \ding{55} & \ding{55} & \Checkmark & \ding{55} & \ding{55} & 4/9 \\
Right to be Forgotten & \Checkmark & \Checkmark & \Checkmark & \ding{55} & \Checkmark & \ding{55} & \ding{55} & \ding{55} & 4/9 \\
Right to Lodge Complaints & \Checkmark & \ding{55} & \Checkmark & \Checkmark & \Checkmark & \ding{55} & \ding{55} & \ding{55} & 4/9 \\
AI-generated Watermarking & \ding{55} & \ding{55} & \ding{55} & \Checkmark & \Checkmark & \Checkmark & \Checkmark & \Checkmark & 5/9 \\
Prompt Guardrail & \ding{55} & \ding{55} & \ding{55} & \ding{55} & \Checkmark & \Checkmark & \Checkmark & \ding{55} & 3/9 \\
Risk Notification & \ding{55} & \ding{55} & \Checkmark & \Checkmark & \Checkmark & \Checkmark & \Checkmark & \ding{55} & 5/9 \\
Data Encryption & \ding{55} & \ding{55} & \Checkmark & \ding{55} & \ding{55} & \Checkmark & \ding{55} & \ding{55} & 2/9 \\
Protection of Minors & \ding{55} & \ding{55} & \Checkmark & \Checkmark & \Checkmark & \ding{55} & \ding{55} & \ding{55} & 3/9 \\ \hline
\end{tabular}
}
\label{table:Tallies}
\end{table*}

%% file: table/Explanations.tex
\begin{table*}[h]
\caption{Explanations about GAI privacy labels. *The GitHub account does not count as a publicly available contact.}
\centering
% \begin{tabularx}{\textwidth}{llll}
\rowcolors{2}{gray!10}{white}
\begin{tabularx}{\textwidth}{>{\hsize=.4\hsize\arraybackslash}X 
    >{\hsize=0.8\hsize\arraybackslash}X 
    >{\hsize=2.2\hsize\arraybackslash}X
    >{\hsize=0.6\hsize\arraybackslash}X
}
\toprule
 & \textbf{Label Field} & \textbf{Explanation} & \textbf{Example Answers}\\ \midrule
Basic Info & Base Model & The names of foundation models that are embedded in this tool. (e.g., GPT-4, GPT-3.5, Ernie, etc) & GPT-3.5/GPT-4/... \\
& Tool Modality & The Modalities of information processed by the reception and response of the tool, respectively. (e.g., text-to-text, image to text) & Text to Image \\
& Tool Functionality & The major capabilities and services provided to users to meet their needs and solve specific problems. & Image Generation \\
 & Working Details & Comprehensive details provided to users about this tool. (e.g., documents about how the system works, data processing process) & A link to the GAI app documentation \\
& Controller Contact & The publicly available contact of the GAI app developers*. (e.g., an email address) & abc@company.com \\
& Target Users & The intended audience or primary user base for this service. & Researchers \\ \midrule\midrule
 & \textbf{Label Field} & \textbf{Explanation} & \textbf{Compliance Status}\\ \midrule
Data Rights & Data Retention & The practice of storing data for a specific period of time. & Yes/No \\
 & Right to Access & The right of users to request to access their collected personal information. & Yes/No \\
 & Right to be Forgotten & The right of users to request to erasure or deletion of their personal information. & Yes/No \\
 & Right to Lodge Complaints & The right of users to lodge a complaint with a supervisory authority. & Yes/No \\ \midrule
Risk Related & AI-generated Watermarking & A machine-readable and detectable mark embedded in content generated or modified by GAI systems. & Yes/No \\
 & Prompt Guardrail & Comprehensive security protocols implemented to scrutinize both user inputs and system outputs for potential malicious activities. (e.g., employing stringent input/output filtering mechanisms) & Yes/No \\
 & Risk Notification & A notification that informs users of the relevant risks they may face when using GAI tools. (e.g. copyright disputes) & Yes/No \\ \midrule
\mbox{Additional Info} & Data Encryption & Data are encrypted and transferred over a secure connection. & Yes/No \\
& Protection of Minors & Special treatment made for the protection and convenience of children. & Yes/No \\ \midrule
\end{tabularx}
\label{table:label explanation}
\end{table*}

%% file: 5_Repo2Label.tex
\begin{figure*}
    \centering
    \includegraphics[width=0.9\textwidth]{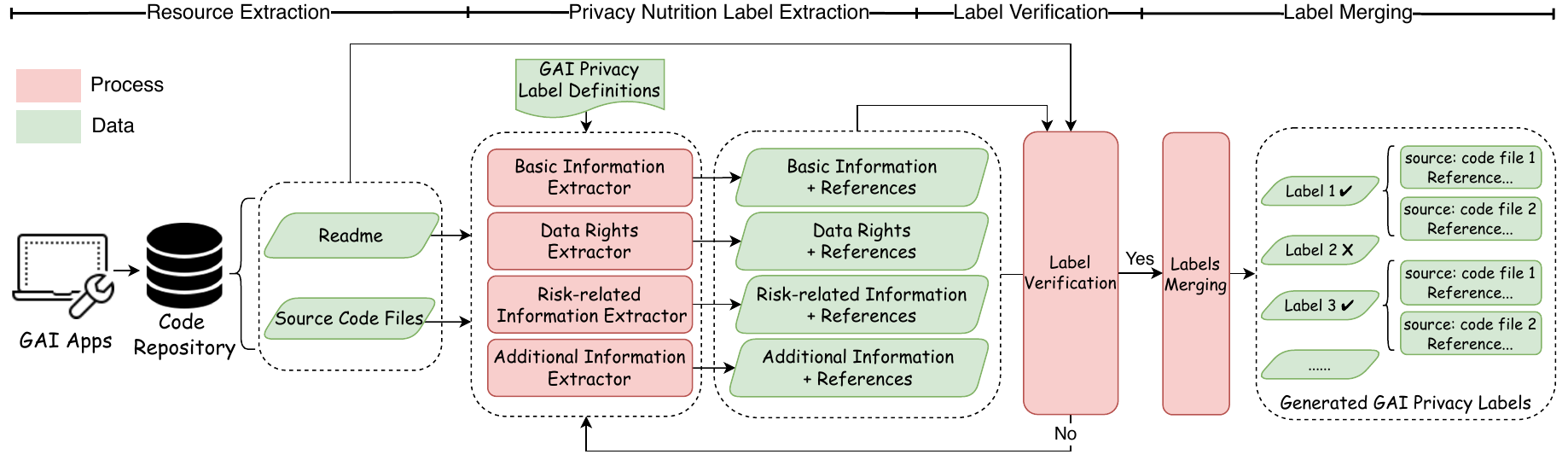}
    \caption{The overview of \texttt{Repo2Label} framework.}
    \label{fig:approach overview}
\end{figure*}

\section{Repo2Label framework}\label{sec:4method}
To response the Challenge-2 discussed in~\autoref{sec_motivation}, we propose \texttt{Repo2Label}, an automated framework that can generate GAI privacy label based on code repository, authentically reflecting the privacy practices of GAI apps.

%给定一个AI Agent，我们的目标是分析它是否存在我们在section 2中提到的nutrition labels。因此我们的framework首先从AI Agent中抽取code repository（包含源代码文件和用来描述代码仓库的readme文件）和external knowledge（比如user guides和product introduction等以文本形式展示的documents）。
%给定每个文件，我们设计了4个AI Uints分别抽取这个文件中的4大类Nutrition Labels。
%为了降低AI Unit导致的幻觉风险，我们通过verify explanation操作check label中可能产生的幻觉，并且通过check结果帮助AI Unit进行反思，进一步抽取准确的label。
%最后，我们merge 所有文件的Nutrition Labels得到Repo-Level的Privacy Nutrition Label。其中，每个label包含对应的source和explanation，帮助用户快速理解这些labels。

\subsection{Overview}
Given a GAI app, our ultimate objective is to analyze its corresponding GitHub repository, determining the value of each GAI privacy label field mentioned in \autoref{sec_label_design}. 
\autoref{fig:approach overview} illustrates the overview of our \texttt{Repo2Label} framework, including four stages.
First, we extract the code repository from a given GAI app for subsequent analysis (\autoref{sec:resource extraction}).
Then, we design multiple AI units to extract information about GAI privacy labels from the code repository and generate explanations for each field (\autoref{sec:label extraction}).
Next, we conduct label verification to assess the accuracy of the provided references and implement reflection processes to correct any hallucinated labels identified by the model (\autoref{sec_label_verification}).
Finally, we combine individually generated GAI privacy labels (for each code file) into a comprehensive repository-level GAI privacy label, with references to further explain the value of each field.(\autoref{sec:label merging}).

\subsection{Resource Extraction} \label{sec:resource extraction}

The only input of the \texttt{Repo2Label} framework is the link of the GitHub repository of the given GAI apps.
Based on the link, we can access all files contained in the repository by the GitHub API~\cite{GitHub_API}.
Normally, these code repositories contain a semi-structured textual document, README, that demonstrate the implementation, and various source code files. 
We exclude irrelevant art files (e.g., images), embedding vectors (e.g., the weights of models), and datasets, since they do not provide any informative hint about the behaviors of GAI apps.
% External knowledge includes user guides and product introductions, offering usage and basic information about the tools.
We filter out those files based on the file type. 
%In the next section, we describe the Privacy Nutrition Label Extraction process to extract privacy nutrition labels from all these files.

\subsection{Privacy Label Extraction} \label{sec:label extraction}
This is the core stage of the whole framework. We aim to analyze the content in the repository to obtain the behaviors that are related to the GAI privacy label.
% Given semi-structured documents (e.g., README) and source code files extracted from \autoref{sec:resource extraction}, our goal is to extract privacy nutrition labels from these resources. 
This task requires understanding the functionality of individual functions within code files and the semantics of natural language descriptions.
Recent studies have shown that large language models (LLMs) are exceptionally capable of software code understanding tasks~\cite{hou2023large, liao2023context, han2024don, pan2024large}.
Considering our need to analyze the semantics of both code and natural language for extracting privacy nutrition label information, we designed several AI units by employing LLMs as our foundational model.
% thanks to their robust in-context learning abilities. 
% These models adeptly adapt to new tasks using precise task descriptions and demonstrations, and are naturally proficient in interpreting natural language texts.

% \begin{figure}%[H]
%     \centering
%     \includegraphics[width=0.48\textwidth]{figure/Nutrition_Label_Unit.pdf}
%     \caption{Nutrition Label Extraction Unit Template. This template can be customized to create various nutrition label extractors by inputting different nutrition labels and their corresponding definitions.}
%     \label{fig:nutrition label unit}
% \end{figure}

\textbf{Privacy Nutrition Label Extraction AI Unit.}
For each AI unit, there are two inputs: a file from the repository and the definition of GAI privacy label fields.
Specifically, all content in the file will be treated as character strings and the definition is the field explanations in~\autoref{table:label explanation}.
We then strategically design the prompt templates to exploit the LLMs' capabilities, enabling the effective extraction of GAI privacy label information from code and textual files.
The prompt template mainly comprises three main components: task description, input, and output. 
% As depicted in Figure~\ref{fig:nutrition label unit}, the \textit{Nutrition Label Extraction Unit Template} comprises three main components: task description, input, and output. 
The task description utilizes a structured prompt to clearly outline the process. 
There are three key components of this prompt as follows.
1) @persona: This component instructs the model to assume the role of an expert data analyzer, ensuring clarity and focus on the task's specific requirements.
2) @terminology: 
Here, essential terms are defined in a keyword format to clarify the context.
For instance, the privacy label field, such as \textit{Base Model} and its corresponding definition, such as \textit{``The names of foundation models that are embedded in this tool. (e.g., GPT-4, GPT-3.5, Ernie, etc)''}
% The same applies for \textit{\#\{Nutrition\_Label\_2\}\#} and \textit{\#\{Nutrition\_Label\_Definition\_2\}\#}, and additional labels and definitions can be incorporated into the template as needed.
3) @instruction: This section furnishes explicit instructions for executing the task:
\begin{itemize} [leftmargin=*]
    \item @command: Offers an overarching strategy for the model to follow, ensuring that the extraction process adheres to defined objectives.
    \item @rule: Outlines the fundamental constraints and considerations that must be observed during the GAI privacy label extraction process.
    \item @Input\_format: Specifies the format of the File Content to be inputted.
    \item @Output\_format: Details the expected presentation format for information extracted from the file content.
\end{itemize}
In the @command, we guide the model through the extraction process step-by-step by employing a chain-of-thought approach, as detailed in the works of Wei et al.\cite{wei2022chain}, Wang et al.\cite{wang2022self}, and Lyu et al.~\cite{lyu2023faithful}.
This structured methodology helps in unfolding the reasoning required for each step of the task, enhancing the model's ability to generate coherent and contextually accurate outputs.
Additionally, in the @rule section, we establish specific guidelines to govern the extraction of nutrition labels by the model, ensuring clarity and accuracy in its tasks:

\begin{itemize} [leftmargin = *]
\item @rule1: Limit the model’s task exclusively to information extraction.
\item @rule2: Require the model to provide a reference for each extraction.
\item @rule3: Ensure that all references provided by the model originate exclusively from the file content. The validity of these references will be evaluated in \autoref{sec_label_verification} to determine if any model-generated hallucinations occur during the extraction process and to enable necessary corrective reflections.
\item @rule4: Reiterate that all references come solely from the file content. Emphasizing this rule multiple times underlines its importance and reinforces the model’s adherence to rigorous standards.
\item @rule5: Reiterate that the model generates results according to the specified output format.
\end{itemize}
In addition to the privacy label content, the second and third rules also fetch the references from the file content that supports the extracted label, serving for the next stage for verification, self-correcting potential hallucinations.
Based on the aforementioned prompt template, we develop four specialized AI units, each designed to target specific sections of the privacy labels:
\begin{itemize} [leftmargin = *]
    \item \textbf{Basic Info Extractor:} This unit utilizes six key labels: \textit{Base Model}, \textit{Tools Modality}, \textit{Tool Functionality}, \textit{Tool's Working Details}, \textit{Controller Contact}, and \textit{Target Users}. It systematically extracts these labels from the input content, each supported by detailed explanations.
    \item \textbf{Data Rights Info Extractor:} Integrating four critical labels: \textit{Data Retention, Right to Access, Right to be Forgotten,} and \textit{Right to Lodge Complaints}. This unit extracts information pertinent to data rights from the input content, ensuring each label is clearly explained.
    \item \textbf{Risk Related Info Extractor:} This unit focuses on extracting three risk-related labels: \textit{AI-generated Watermarking, Prompt Guardrail}, and \textit{Risk Notification}. Each label is detailed within the input to provide a clear contextual understanding.
    \item \textbf{Additional Info Extractor:} Targeting additional factors, this unit processes two labels: \textit{Data Encryption} and \textit{Protection of Minors}. 
\end{itemize}
% Due to the page limitation, all prompts for AI units are available in the ``prompts/'' folder of our artifact package~\cite{our_repo}.

\subsection{Label Verification} \label{sec_label_verification}

% Why do we need the verification? LLM Hallucination is widely criticized. 
% LLM no hallucination AI model -$>$ wrong answer.
% repo2label (if there is something wrong, the results are bad)
% -$>$ so we need the verification to double-check/scrutinize the initial results

LLMs are widely criticized for their tendency to produce hallucinations~\cite{ji2023survey, huang2023survey, zhao2023survey}. 
This phenomenon occurs because the inherently probabilistic nature of LLMs can result in outputs that appear convincingly accurate but are actually incorrect, making it difficult for humans to discern the inaccuracies. 
Given this, privacy labels, which serve as crucial privacy notices for GAI app users, must strive to be as accurate as possible.
Studies also show that LLMs can self-correct the hallucinations by proper strategies~\cite{pan2023automatically}.
Therefore, we design the \textit{Label Verification} to double-check the potentially incorrect extracted privacy labels.
Specifically, instead of directly checking the labels, we employ the AI unit to review the correctness of the references used by the model during the label-generation process.
We use the string matching techniques to confirm whether the reference indeed exists within the file content.
If the reference is indeed in the file content, we consider the corresponding label generated by the model to be likely accurate and thus retain it.
Conversely, if the reference is incorrect (i.e., not exist in the original file), we ask the AI unit to proceed with the reflection prompt to regenerate the label with the specific instruction:
\textit{``You previously extracted a label with an incorrect reference that does not exist in the file content. Please ensure that the reference provided this time is present in the file content.''}
If the model provides incorrect references more than three times for the same label, we categorize the label as \textit{N/A}, indicating its non-applicability or unreliability.

\subsection{Labels Merging} \label{sec:label merging}
After the previous stages, we obtain all extracted GAI privacy labels and references from each file.
We then aggregate them as a complete GAI privacy label based on the whole repository.
For each nutrition label, from a set of \textit{n} files, we collect all labels that are not N/A. 
We record the source code path and the corresponding reference for each valid label.
The format used for this recording is $<$file\_path, nutrition\_label, reference$>$. For example,
\begin{itemize}[leftmargin=*]
\item codeRepos/babyagi/babyagi.py
\item gpt-3.5-turbo
\item LLM\_MODEL = os.getenv(`LLM\_MODEL',os.getenv (`OPENAI\_API\_MODEL', `gpt-3.5-turbo')).lower()
\end{itemize}
To generate a repository-level nutrition label, we merge the labels from all files, taking their union. 
For instance, if the base model label across different files includes versions like \textit{gpt-4}, \textit{gpt-3.5-turbo}, and \textit{text-embedding-ada-002}, the final label for the repository would be the combination of these three.
Additionally, the final repository-level nutrition label features an expansion window that allows users to view all the references for each label.
This feature provides a transparent view of the evidence supporting each label, enhancing the credibility and usability of our extracted information.

%% file: 6_Evaluation.tex
\section{Evaluation}\label{sec:5evaluation}
In this section, we systematically evaluate 1) the design of the proposed GAI privacy label format and 2) the performance of \texttt{Repo2Label} framework.

% \xiaoyu{Please clearly specify which research questions are addressed in the evaluation section}

% \xiaoyu{The current research questions are insufficient, and there are many results for Repo2label that have not been reported. For example, the following aspects should at least be reported:
% 1. Readability: Conduct a user study to demonstrate whether Repo2label can really help users quickly and accurately understand the privacy policy.
% 2. Provide some cases where GPT misclassifies to show the actual impact on users when GPT does not produce correct results.
% 3. Previously mentioned, 12\% of GAI repos have a privacy policy. Does Repo2label's result differ from this, and if so, by how much? Are there any errors in the manually constructed privacy policy? Are there any errors in the labels generated by Repo2label? How accurately does Repo2label identify human\-written privacy policies?
% 4. To verify the effectiveness of Repo2label, conduct another user study by having developers use this tool, and then measure the tool's usefulness through their feedback.}

\subsection{Evaluation of GAI Privacy Label Design}\label{sec_eva_label_design}

% refer to： “Okay, whatever”: An Evaluation of Cookie Consent Interfaces
To examine the design of our proposed GAI privacy label format, we conducted a comprehensive online survey as a human evaluation.
Ethical approval for this research was secured from our institution’s Institutional Review Board (IRB). 
It is important to note that before the formal experiment, we conducted a small-scale pilot study.
This pilot study allowed us to preliminarily assess the duration required for the formal study, which averaged 11 minutes and 53 seconds. More critically, through actual task performance and participant feedback, we identified and refined ambiguous statements within the scales.

\subsubsection{Participants Recruitment} Following the research approach proposed by Lin et al.~\cite{lin2023data}, we opted to recruit participants via the crowd worker platform Prolific\footnote{\href{https://www.prolific.com}{https://www.prolific.com}}. Recognizing the findings of Hasegawa et al.~\cite{hasegawaweird} regarding the demographics of participants in Usable Privacy and Security (UPS) field studies, which often exhibit a strong bias towards individuals from WEIRD (Western, Educated, Industrialized, Rich, and Democratic) countries. To proactively avoid the potential unrepresentativeness in our experimental results, we employed a purposive sampling method ~\cite{tongco2007purposive} to balance the distribution of participants across gender, age, and nationality, thereby ensuring diversity in these aspects among our participants.
With a total planned recruitment of 48 participants, we divided the experiment into two stages. Initially, we released 20 slots on the Prolific platform. Subsequently, we adjusted our recruitment criteria (age, nationality) to facilitate the participation of individuals who had not yet taken part in our experiment. 
Finally, we recruited 22 females and 26 males. 20 of them were from developing countries (normally regarded as non-WEIRD) and 28 were from developing countries (normally regarded as WEIRD).

\subsubsection{Survey Design}
We conducted the online survey via the Prolific platform to evaluate participants' assessments of the design of the GAI privacy labels. 
After presenting the informed consent to the participants, we provided a detailed information sheet as an introduction to the concept of the GAI privacy labels.
We also listed several examples to help participants gain a better understanding.
Our evaluation questionnaire primarily focused on key dimensions such as the understandability and interpretability of the GAI privacy labels, and participants were asked to rate statements (5-point Likert scale) about those aspects.
In response to previous UPS studies about the issues of privacy notices, including excessive terminology and lengthy content ~\cite{krumay2020readability, kincaid1975derivation}, poor practical utility because of unacceptable reading costs ~\cite{mcdonald2008cost}, and the importance of standardized privacy notification formats~\cite{kelley2010standardizing}, therefore; we paid particular attention to aspects on user needs, reading burden, and concise representation to ensure the GAI privacy label is both normative and practical. 
Detailed questions are displayed in \autoref{tab_eval_human_result}.

\input{table/evaluation_result}

\subsubsection{Results Analysis}
\autoref{tab_eval_human_result} shows the results of the human evaluation.
Overall, the proposed design of GAI privacy labels received high ratings across all eight dimensions. In particular, \textit{Concise Representation} and \textit{Appropriate Volume} were rated the highest, with over 80\% of participants selecting ``Agree'' even ``Strongly Agree''. 
This indicates that the GAI privacy label effectively delivers information in a succinct manner without being redundant, thus meeting the informational needs of users efficiently. 
Conversely, \textit{Reading Burden} was identified as the dimension with relatively lower agreement. 
Nevertheless, 71\% of participants still rated this dimension as ``Agree'' or ``Strongly Agree''. The higher mean (4.02) and median (4) scores for this dimension further support this finding, suggesting that while there is room for improvement, the overall user perception is that the reading burden remains within acceptable limits.
Overall, the proposed design of GAI privacy labels receives greatly positive ratings across multiple critical aspects. These findings highlight the label's quality and high level of user endorsement.

\subsection{Evaluation of \texttt{Repo2Label} Framework}
The accuracy of privacy labels is crucial, as inaccuracies can lead to information overload for users and severely undermine their trust in the software. 

\subsubsection{Dataset Construction}~\label{sec_data_construction}
In \autoref{sec_market_app_analysis}, we have identified 148 GAI apps and their GitHub repositories. 
Given the absence of ground truth for privacy labels for GAI apps, we manually crafted a benchmark dataset for the evaluation.
As this is a time-consuming task, we randomly sampled approximately 20\% of the repositories, resulting in a total of 29 repositories for manual annotation. This involved 922 code files, with an average of 160 lines of code per file. 
Two experienced researchers painstakingly examined the content within the repository files of sampled GAI apps, and manually annotated the values of their privacy label field according to the current file.
Both annotators have at least three years of experience in AI4SE (AI for Software Engineering) research and two years in UPS research.
~\autoref{lst:Right to be Forgotten} presents an example where the label \textit{Right to be Forgotten} is marked as `Yes'. In this case, the GAI app offers users the functionality to clear their conversation history. 
Similarly,~\autoref{lst:AI-generated Watermarking} illustrates a code example where the label field \textit{AI-generated Watermarking} is marked as `Yes'. In this instance, the GAI app is capable of adding watermarks to generated images, thereby mitigating potential copyright issues. 
Furthermore, when functions like the one depicted in \autoref{lst:Prompt Guardrail}, which perform security checks on user inputs, are present in the repository, the \textit{Prompt Guardrail} label is accordingly marked as `Yes'. Examples of \textit{Right to Lodge Complaints}, \textit{Risk Notification} and the \textit{Data Encryption} are illustrated in \autoref{lst:Right to Lodge Complaints}, \autoref{lst:Risk Notification} and \autoref{lst:Data Encryption}, respectively.

\begin{lstlisting}[language=Python,caption={A code example of \textit{Right to be Forgotten} label from \href{https://github.com/clmnin/summarize.site}{Large Language and Vision Assistant} (699 stars).}, label={lst:Right to be Forgotten}]
async function deleteConversation(conversationId) {
  const accessToken = await getAccessToken();
  const resp = await fetch(
    `https://chat.openai.com/backend-api/conversation/${conversationId}`,
    {
      method: "PATCH",
      headers: {
        "Content-Type": "application/json",
        Authorization: `Bearer ${accessToken}`,
      },
      body: JSON.stringify({ is_visible: false }),
    }
  )
    .then((r) => r.json())
    .catch(() => ({}));
  if (resp?.success) {
    return true;
  }
  return false;
}
\end{lstlisting}

\begin{lstlisting}[language=Python,caption={A Code example of \textit{AI-generated Watermarking} label from \href{https://github.com/CompVis/stable-diffusion}{Stable Diffusion} (63k stars).}, label={lst:AI-generated Watermarking}]
print("Creating invisible watermark encoder (see https://github.com/ShieldMnt/invisible-watermark)...")
    wm = "StableDiffusionV1"
    wm_encoder = WatermarkEncoder()
    wm_encoder.set_watermark('bytes', wm.encode('utf-8'))
\end{lstlisting}
\vspace{-5pt}

\begin{lstlisting}[language=Python,caption={A code example of \textit{Prompt Guardrail} label from \href{https://github.com/run-llama/llama_index}{GPT Index} (27k stars).}, label={lst:Prompt Guardrail}]
def from_defaults(
    cls,
    temperature: float = 0.7,
    answer_style: int = 1,
    safety_setting: List["genai.SafetySetting"] = [],
) -> "GoogleTextSynthesizer":
    """Create a new Google AQA.

    Example:
      responder = GoogleTextSynthesizer.create(
          temperature=0.7,
          answer_style=AnswerStyle.ABSTRACTIVE,
          safety_setting=[
              SafetySetting(
                  category=HARM_CATEGORY_SEXUALLY_EXPLICIT,
                  threshold=HarmBlockThreshold.BLOCK_LOW_AND_ABOVE,
              ),
          ]
      )......
\end{lstlisting}

\begin{lstlisting}[caption={A code example of \textit{Right to Lodge Complaints} label from \href{https://github.com/haotian-liu/LLaVA}{Large Language and Vision Assistant} (13k stars).}, label={lst:Right to Lodge Complaints}]
Please click the "Flag" button if you get any inappropriate answers! We will collect those to keep improving our moderator.
\end{lstlisting}

\begin{lstlisting}[caption={A code example of \textit{Risk Notification} label from \href{https://github.com/allenai/macaw}{Macaw} (456 stars).}, label={lst:Risk Notification}]
## Disclaimer

As a model capable of generating free form text, the output of the model is not guaranteed to be free of
offensive material, so appropriate caution is advised when using the model.
\end{lstlisting}

\begin{lstlisting}[caption={A code example of \textit{Data Encryption} label from \href{https://github.com/reworkd/AgentGPT}{AgentGPT} (29k stars).}, label={lst:Data Encryption}]
def test_encrypt_decrypt():
    key = Fernet.generate_key()
    service = EncryptionService(key)

    original_text = "Hello, world!"
    encrypted = service.encrypt(original_text)
    decrypted = service.decrypt(encrypted)

    assert original_text == decrypted
\end{lstlisting}

Ultimately, 13,830 labels were annotated based on 922 code files in 29 repositories. It took an average of 5 minutes to annotate each code file.
For any disagreement in the annotation, they discussed and agreed on the same answer, and if the
disagreement persisted, a senior researcher joined the discussion to facilitate a resolution.
The Cohen's Kappa of initial annotation is $\kappa=0.78$, indicating a high level of inter-rater agreement.

\begin{figure}%[H]
    \centering
    \includegraphics[width=0.48\textwidth]{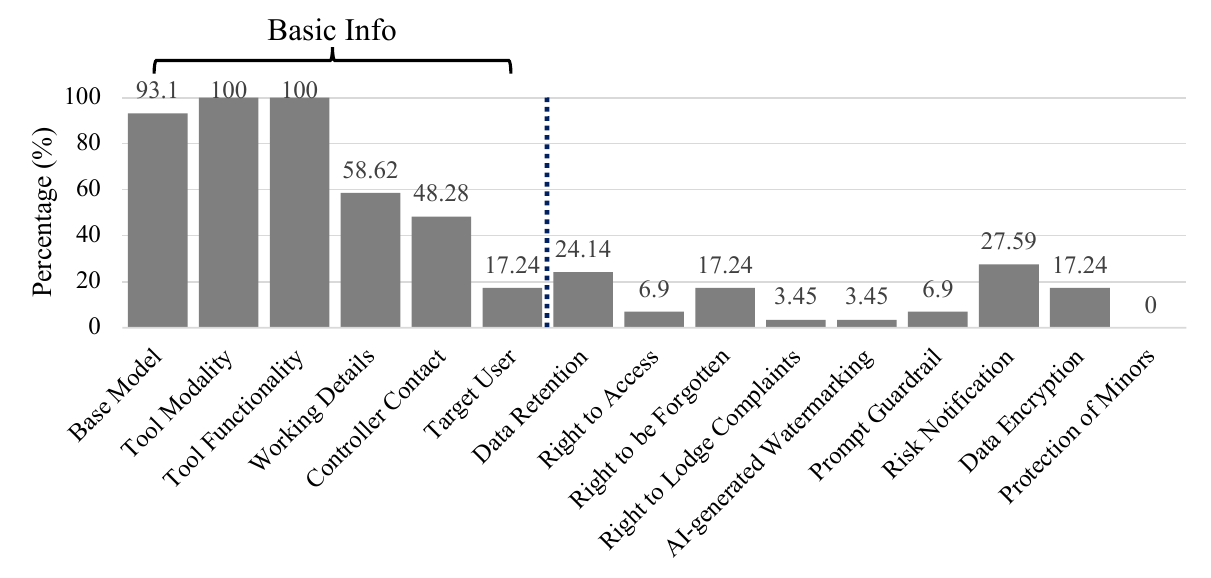}
    \caption{Frequency of the GAI Privacy Label fields marked as ``Yes'' in the manually annotated dataset.}
    \label{fig:percentage of nutrition label}

\end{figure}

\autoref{fig:percentage of nutrition label} shows the frequency of the GAI Privacy Label fields marked as ``Yes''. All 29 repositories include information on \textit{Tool Modality} and \textit{Tool Functionality}. 
It is important to note that not every repository explicitly specifies the base model used, as sometimes this can be determined by the user input and some of the apps are Visual Studio Code plugins. 
Additionally, \textit{Working Details} and \textit{Controller Contact} exist in about half of the examined repositories.
In the Basic Info section, the ratio of \textit{target user} is the lowest. 
Despite this attention to \textit{Basic Info}, we observed a notable deficiency in other sections. 
Only \textit{Data Retention}, \textit{Risk Notification}, \textit{Right to Access}, and \textit{Right to Be Forgotten} are sporadically covered, with each category appearing infrequently. 
Moreover, none of the items within \textit{Protection of Minors} were detected in any real-world apps.
These observations suggest that while developers consistently focus on basic information, they often overlook other crucial privacy-related alignment measurements during the development process, potentially leaving significant gaps in privacy considerations. A comprehensive analysis of these findings is provided in \autoref{sec_discussion_incompliance}.

\input{table/performance}

\subsubsection{Evaluation of \texttt{Repo2Label}}
% To assess the effectiveness of \texttt{Repo2Label} in extracting privacy nutrition labels, we input 922 code files from \autoref{sec_data_construction} into our framework, resulting in 15 predicted labels per file, totaling 13,830 predicted labels.
% Due to our focus on maximizing the detection of privacy labels within repositories, we prioritized recall over precision.
\autoref{table:performance} demonstrates the performance of \texttt{Repo2Label} in generating GAI privacy labels. 
We utilized widely recognized and high-performing foundation LLMs, GPT-4o~\cite{OpenAI_model} and GPT-4 Turbo~\cite{OpenAI_model}, to drive the AI units in the framework. 
Additionally, we conducted a comparative analysis of the results before and after applying the verification process described in \autoref{sec_label_verification}.
Given the In-Context Learning (ICL) capabilities of LLM~\cite{dong2022survey}, we also compared results under zero-shot and few-shot settings.

Results show that GPT-4o greatly outperforms GPT-4 Turbo under all settings, achieving a 0.84 F1-score under the optimal settings, compared to a 0.64 F1-score. 
Contrary to our expectations, the few-shot strategy does not significantly increase the performance compared to the zero-shot, especially for GPT-4 Turbo.
Upon manual inspection of the result from \texttt{Repo2Label}, we found that not providing sufficiently representative examples in the few-shot learning could cause the LLM to capture irrelevant content.
For instance, when given examples related to the \textit{Protection of Minors}, as illustrated in \autoref{lst:Protection of Minors}, the LLMs incorrectly classified all code files dealing with Sexual Content as related to the \textit{Protection of Minors}.
This misclassification occurred because the provided examples led the LLMs to generalize the context overly.

Additionally, the verification step led to improvements in all metrics. 
Specifically, with the GPT-4o model and a zero-shot setup, the app of verification resulted in precision increasing from 0.68 to 0.81 (+19.12\%), recall from 0.84 to 0.88 (+5.95\%), and F1-score from 0.75 to 0.84 (+13.33\%). 
Other configurations involving few-shot learning and GPT-4 Turbo also showed improvements after verification, although the extent varied. The least improvement was observed with the GPT-4 Turbo in a zero-shot setup.
In short, using GPT-4o with the zero-shot method followed by verification achieved the best performance, with a precision of 0.81, recall of 0.88, and F1-score of 0.84.

\begin{lstlisting}[caption={A code example of \textit{Protection of Minors} label from \href{https://github.com/run-llama/llama_index}{GPT Index} (27k stars).}, label={lst:Protection of Minors}]
Llama Guard safety taxonomy:

- Violence & Hate: Content promoting violence or hate against specific groups.
- Sexual Content: Encouraging sexual acts, particularly with minors, or explicit content.
\end{lstlisting}

% It is noteworthy that, during our experiments, we did not extract label information from code files exceeding 8000 tokens to conserve token resources. This constraint significantly contributes to the discrepancies between the labels generated by the \texttt{Repo2Label} framework and the ground truth.
% However, we believe that increasing the token limit threshold and more powerful foundation LLMs could further enhance the performance of \texttt{Repo2Label}.

\subsubsection{Comparison between Repo2Label and self-declared privacy policies}
\input{table/compare_pp_label}

Among the 29 annotated GAI apps and their repositories, 11 GAI apps provide a privacy policy.
We then manually scrutinize their privacy policies in terms of the aspects covered by the GAI privacy labels, compared against the benchmark annotations.
\autoref{compare_pp_label} presents the results of the disclosures in GAI app privacy policies, with an overall F1-score of only 0.13.
This indicates that the long-standing issue of under-disclosure in privacy policies also persists in the GAI app context, and may even be more pronounced.
Also, as lack of enforcement force from a central market, such as the Google Play app store for mobile apps, developers do not proactively work on providing authentic privacy policies. For example, \href{https://github.com/haotian-liu/LLaVA}{Large Language and Vision Assistant} has attracted over 13k GitHub stars, but their privacy policy\footnote{\href{https://domeccleston.notion.site/Privacy-Policy-73a5302dc77349a6892d632d975582bf}{Privacy Policy of Large Language and Vision Assistant}} only contains 63 words.
Despite the statements of various data rights in examined privacy policies, our manual annotation results reveal that the actual implementation of these apps does not align with their stated privacy policies. For instance, 72.2\% of the privacy policies claim that their services are not intended for use by children. However, there is no implementation of age verification or other validation functions in the code to implement this disclosure.

%% file: table/evaluation_result.tex
\begin{table*}[t]
\centering
\caption{Questionnaire and results of human evaluation for our GAI privacy label design. ``Mdn.'' stands for Median. ``Distr.'' stands for the Distribution of the responses (from left to right: strongly disagree to strongly agree).}
\label{tab_eval_human_result}
\resizebox{1.0\textwidth}{!}{%
\begin{tabular}{l|c|l|c|c|c|c}
\toprule

\textbf{No.} & \textbf{Aspect} & \textbf{Question} & \textbf{Mean} & \textbf{Mdn} & \textbf{SD} & \textbf{Distr.} \\

% \midrule

% \textbf{No.} & \textbf{Aspect} & \textbf{Question} & \textbf{Mean} & \textbf{Mdn} & \textbf{SD} & \textbf{Distr.} \\

% \midrule
% \multicolumn{6}{c}{\textbf{Demographics Questions}} \\
% \midrule

% $\textit{D}_1$  &\makecell[l]{What gender do you identify with?} & 3.91 & \raisebox{-0.32\totalheight}{
% \includegraphics[width=0.04\textwidth]{Figures/question_1_bar_chart.png}
% } 

% \\

\midrule

$\textit{Q}_1$  & \textbf{User Needs} &\makecell[l]{The GAI privacy nutrition label provides the information I concern about\\ in terms of online privacy.} & 3.98 & 4 & 1.11 & \raisebox{-0.32\totalheight}{\includegraphics[width=0.04\textwidth]{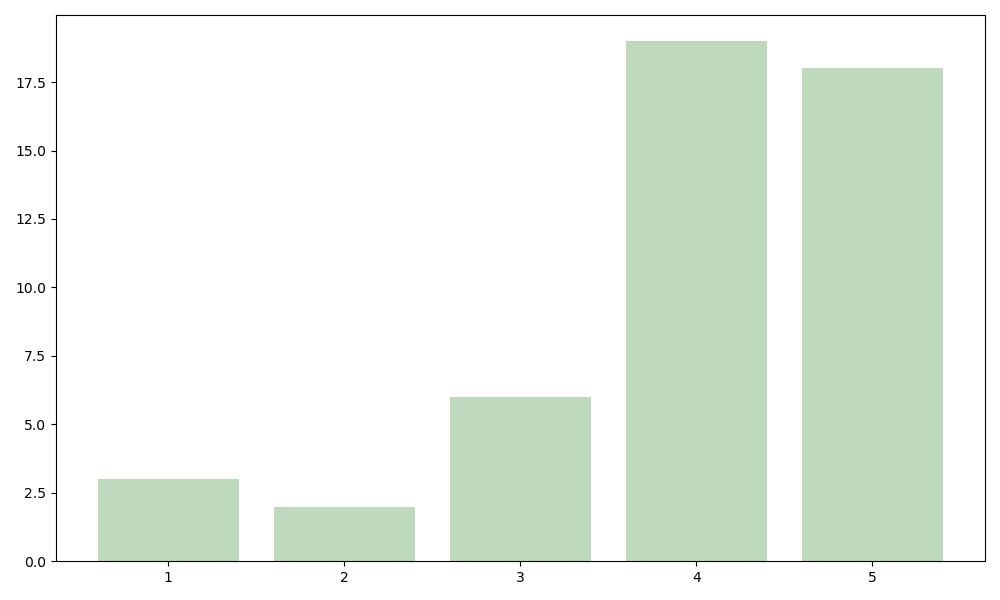}} \\

$\textit{Q}_2$ & \textbf{Reading Burden}  &\makecell[l]{The GAI privacy nutrition label can relieve my reading pressure compared\\ to normal privacy policies.} & 4.02 & 4 & 1.11 & \raisebox{-0.32\totalheight}{\includegraphics[width=0.04\textwidth]{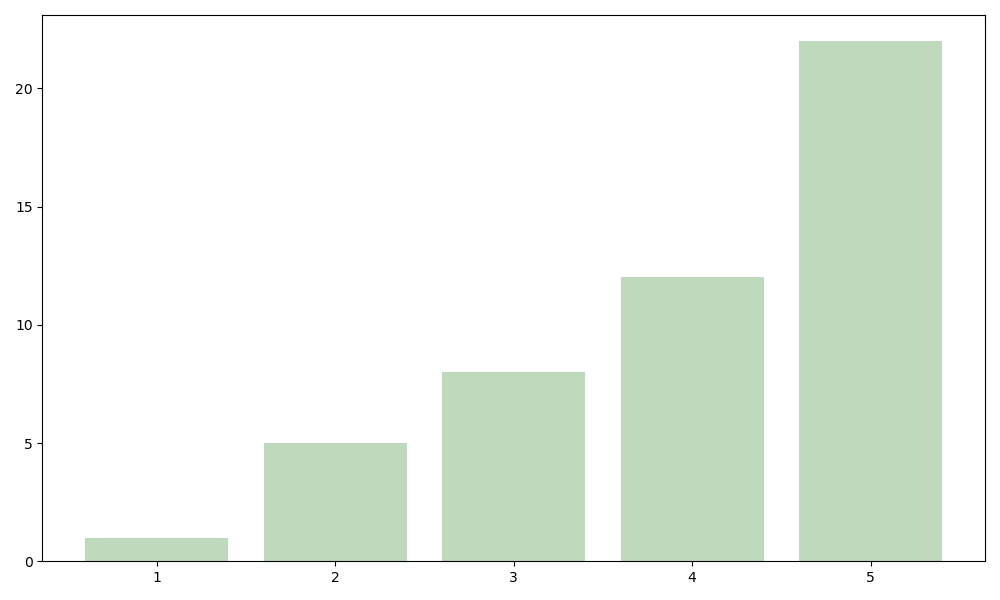}} \\

$\textit{Q}_3$ & \textbf{Privacy Assurance} &\makecell[l]{The GAI privacy nutrition label can enhance privacy assurance.} & 4.08 & 4.5 & 1.20 & \raisebox{-0.32\totalheight}{\includegraphics[width=0.04\textwidth]{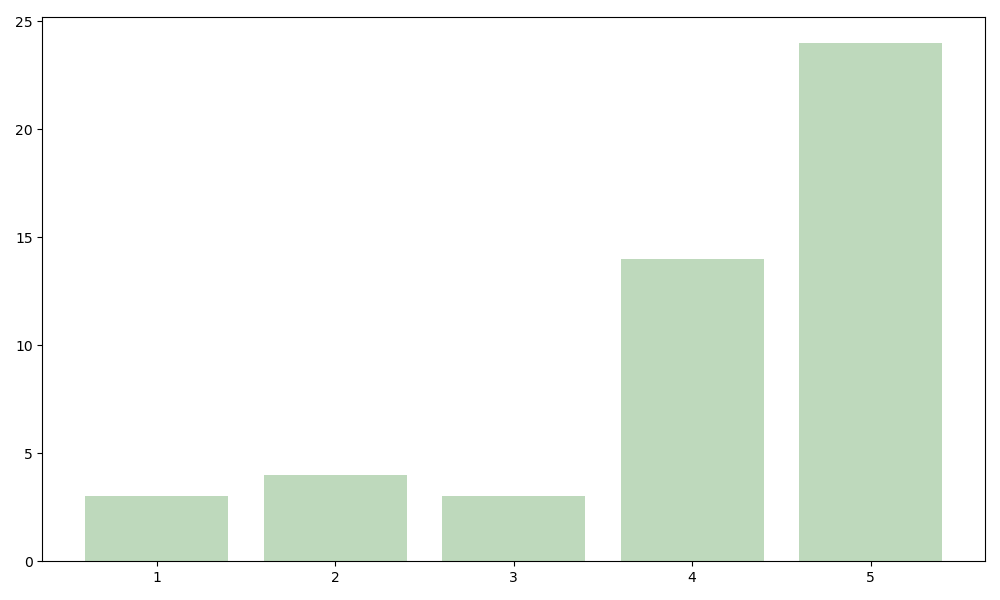}} \\

$\textit{Q}_4$ & \textbf{Trackability}  &\makecell[l]{For each label data field, the GAI privacy nutrition label can provide the\\ source and the reason.} & 4.21 & 5 & 1.12 & \raisebox{-0.32\totalheight}{\includegraphics[width=0.04\textwidth]{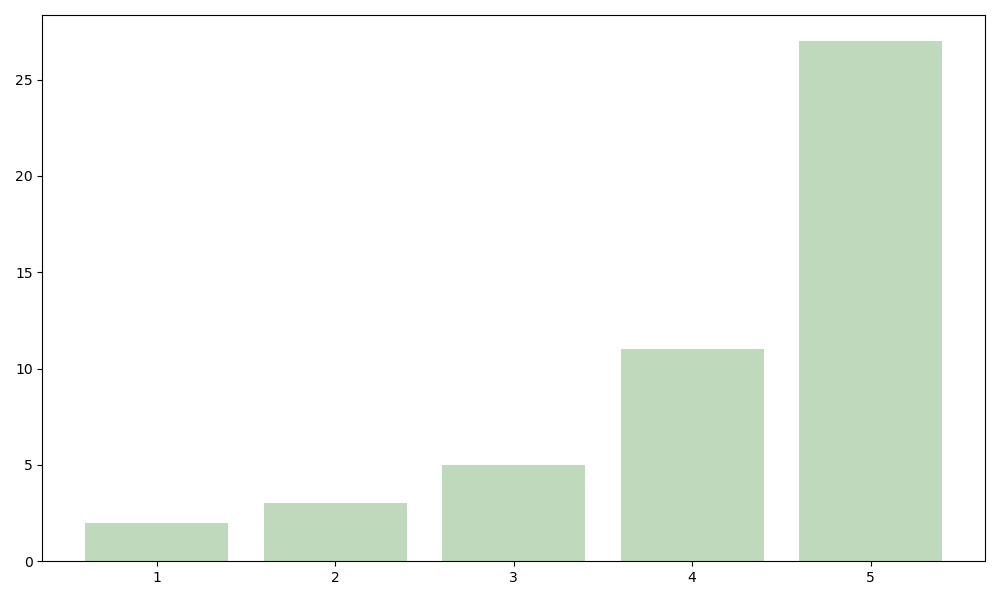}} \\

$\textit{Q}_5$ & \textbf{Understandability} &\makecell[l]{The GAI privacy nutrition label is easy to be comprehended in a timely\\ manner.} & 4.06 & 4 & 1.03 & \raisebox{-0.32\totalheight}{\includegraphics[width=0.04\textwidth]{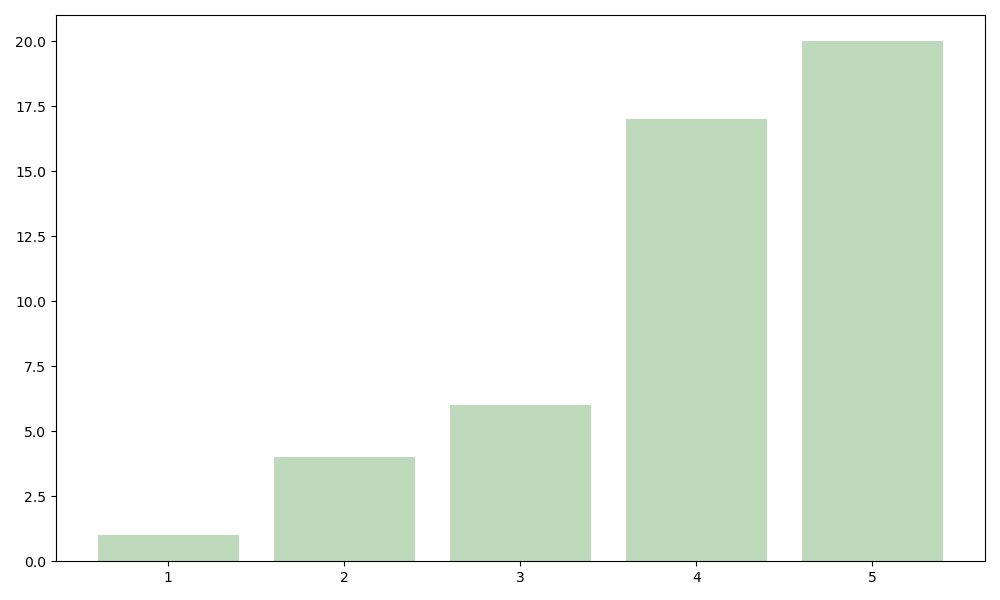}} \\

$\textit{Q}_6$ & \textbf{Interpretability} &\makecell[l]{It is easy to interpret the meaning of each data field of the GAI privacy\\ nutrition label.} & 4.31 & 5 & 0.98 & \raisebox{-0.32\totalheight}{\includegraphics[width=0.04\textwidth]{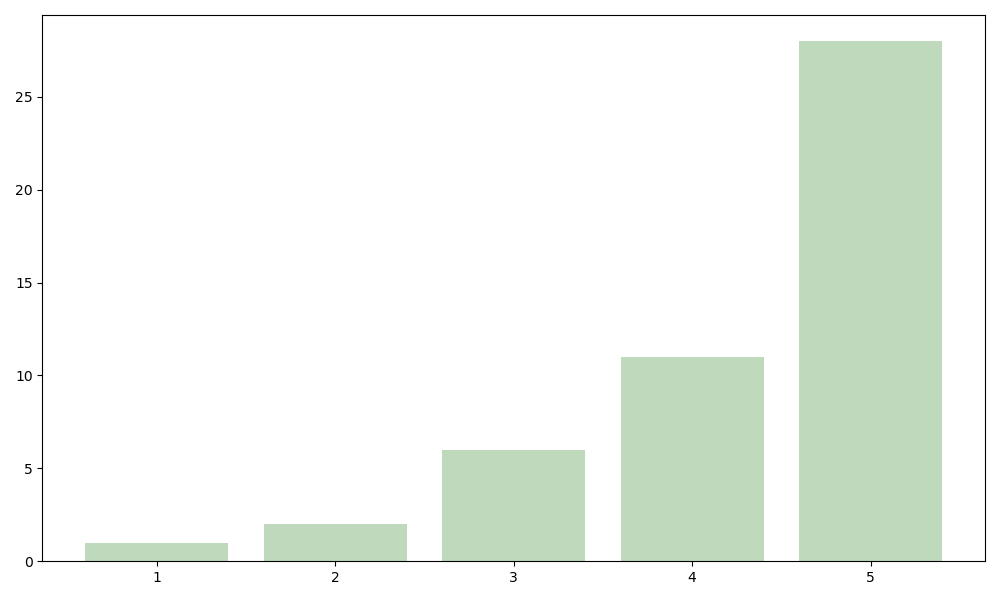}} \\

$\textit{Q}_7$ & \textbf{Concise Representation}  &\makecell[l]{The representation and design of the GAI privacy nutrition label are \\compact and concise.} & 4.33 & 5 & 1.12 & \raisebox{-0.32\totalheight}{\includegraphics[width=0.04\textwidth]{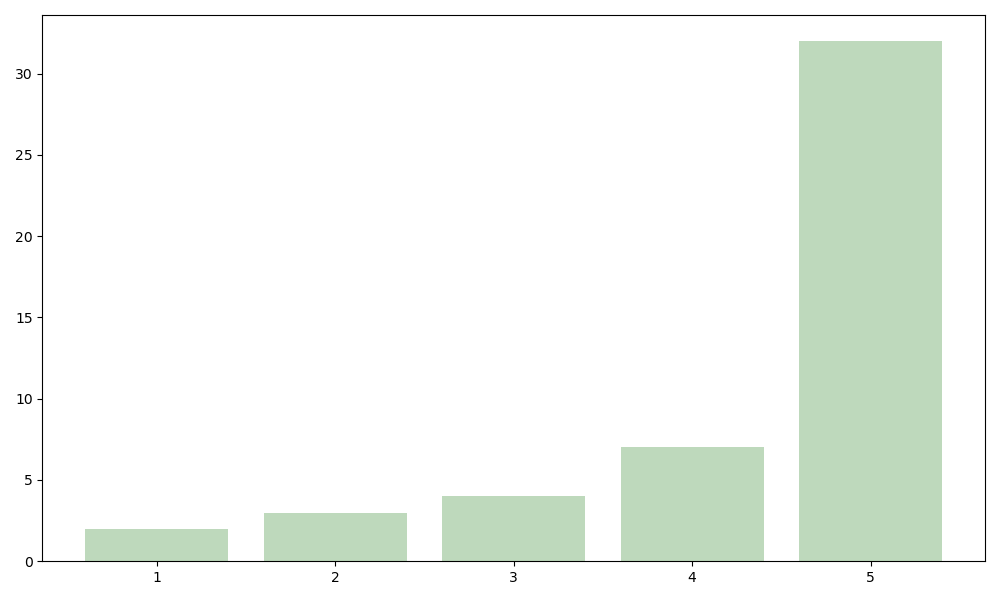}} \\

$\textit{Q}_8$  & \textbf{Appropriate Volume} &\makecell[l]{The volume of GAI privacy nutrition label is neither too much nor too little,\\ and it can be read in a timely manner.} & 4.5 & 5 & 0.89 & \raisebox{-0.32\totalheight}{\includegraphics[width=0.04\textwidth]{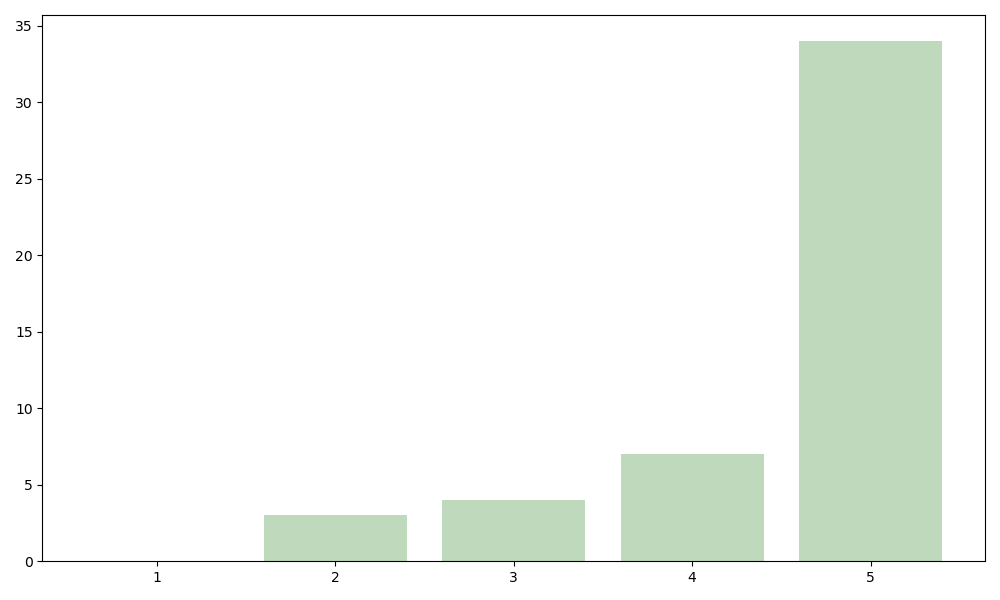}} \\

\bottomrule
\end{tabular}
}%
\vspace{-10pt}
\end{table*}

%% file: table/performance.tex
\begin{table*}
\centering
\caption{Performance of Repo2Label framework. ``Prec.'' stands for precision and ``Rec.'' stands for recall. ``V'' stands for the verification stage. GPT-4 Turbo is a large multimodal model released by OpenAI.
GPT-4o (``o'' for ``omni'') is the most advanced, multimodal flagship model released in May 2024. The GPT-4o is twice as fast as the GPT4-turbo, 50\% cheaper, and significantly better at handling text in non-English languages~\cite{OpenAI_model}.}
\begin{tabular}{ll|lll|lll|lll|lll|lll}
 \toprule
 &  & \multicolumn{3}{c|}{\textbf{Basic Info}} & \multicolumn{3}{c|}{\textbf{Data Rights}} & \multicolumn{3}{c|}{\textbf{Risk Related}} & \multicolumn{3}{c|}{\textbf{Additional Info}} & \multicolumn{3}{c}{\textbf{Average}} \\ \midrule
\textbf{LLM} & \textbf{Settings} & \textbf{Prec.} & \textbf{Rec.} & \textbf{F1} & \textbf{Prec.} & \textbf{Rec.} & \textbf{F1} & \textbf{Prec.} & \textbf{Rec.} & \textbf{F1} & \textbf{Prec.} & \textbf{Rec.} & \textbf{F1} & \textbf{Prec.} & Rec. & \textbf{F1} \\ \midrule
GPT-4o & zero-shot & .68 & .84 & .75 & \textbf{.76} & \textbf{.73} & \textbf{.74} & .78 & \textbf{.90} & .84 & .62 & \textbf{.83} & \textbf{.71} & .68 & .84 & .75 \\
 & zero-shot + V & \textbf{.81} & \textbf{.89} & \textbf{.85} & \textbf{.76} & .70 & .73 & \textbf{.86} & \textbf{.90} & \textbf{.88} & \textbf{.83} & .62 & \textbf{.71} & \textbf{\uline{.81}} & \textbf{\uline{.88}} & \textbf{\uline{.84}} \\
 & few-shot & .67 & .83 & .74 & \textbf{.76} & \textbf{.73} & .74 & .50 & .89 & .64 & .62 & \textbf{.83} & \textbf{.71} & .67 & .83 & .74 \\
 & few-shot + V & .80 & .87 & .84 & \textbf{.76} & .70 & .73 & .49 & .89 & .63 & .62 & \textbf{.83} & \textbf{.71} & .79 & .87 & .83 \\ \midrule
GPT-4 Turbo & zero-shot & .67 & .69 & .68 & .50 & .19 & .27 & .42 & .84 & .56 & .23 & .50 & .32 & .64 & .68 & .66 \\
 & zero-shot + V & .69 & .70 & .70 & .50 & .19 & .27 & .57 & .84 & .68 & .25 & .50 & .33 & .68 & .69 & .68 \\
 & few-shot & .58 & .72 & .64 & .59 & .63 & .61 & .61 & .74 & .67 & .18 & .43 & .25 & .58 & .71 & .64 \\
 & few-shot + V & .60 & .73 & .66 & .58 & .70 & .63 & .55 & .84 & .67 & .38 & .50 & .43 & .60 & .73 & .66 \\ \midrule
\end{tabular}
\label{table:performance}
\end{table*}

%% file: table/compare_pp_label.tex
\begin{table}[]
\centering

\caption{Results of quality examination on self-declared privacy policies of GAI apps.}
\begin{tabular}{llll}
\toprule
 & \textbf{Precision} & \textbf{Recall} & \textbf{F1} \\
\midrule
Basic Info & 0.17 & 0.02 & 0.04 \\
Data Rights& 0.12 & 0.38 & 0.18 \\
Risk Relate & 0.50 & 0.25 & 0.33 \\
Additional Info & 0.10 & 0.20 & 0.13 \\
\midrule
Overall  & 0.15 & 0.11 & 0.13 \\ \bottomrule
\end{tabular}
\label{compare_pp_label}
\end{table}

%% file: 7_Discussion.tex
\section{Discussion}\label{sec:6discussion}
This discussion aims to provide a reference and basis for a more comprehensive understanding of the significance and implications of our study.
% refer to shidong's paper:https://www.usenix.org/system/files/sec24fall-prepub-303-pan-shidong-hope.pdf#page=13.22

\subsection{Regulatory (In)compliance of Open-source GAI Apps}~\label{sec_discussion_incompliance}
By observing the manually annotated dataset of privacy practices of GAI apps, we notice a significant discrepancy between GAI app implementation and regulations.
Notably, a label marked as ``No'' does not necessarily indicate that the GAI app violates regulations, and vice versa. 
There are several reasons for this: 1) Open-source GAI apps are not commercial software, and they may not be subject to certain regulations; 2) Regulations vary by region, and their interpretation could be subjective and controversial; and 3) The implementation of privacy practices may not rigorously comply against regulatory requirements.
Nonetheless, our findings still reflect that the open-source GAI apps community commonly lacks awareness and responsiveness to those regulations.
At the early stage of the market, compliance often cannot be guaranteed, but we advocate for the community to pay attention to these requirements.
% In addition, those GAI apps have a high popularity 
% This oversight can lead to significant privacy risks for users and potential legal repercussions for developers.

\subsection{Comparison with Other Privacy Notices}
% for expert versus non-expert
To establish transparency surrounding AI-enabled products (e.g., GAI apps) and thus yield trust, researchers have also proposed various types of privacy notices in addition to privacy policies and privacy labels, such as the Model Cards~\cite{mitchell2019model, liang2024s} and AI Bills of Materials (AIBOMs)~\cite{xia2023trust}. 
Model cards are files that accompany the foundation models (e.g., Llama) and provide handy information. They are essential for discoverability, reproducibility, and sharing.
Typically, Model Cards coefficiently provide critical information about the model's dataset, evaluation results, and potential ethical considerations.
However, Model Cards do not provide privacy disclosures and the dynamic nature of model development presents challenges in maintaining the model card content timely.
AIBOMs offers a more specialized disclosure aimed at experts, focusing on elements such as software dependency, version \& licenses, etc. 
While model cards and AIBOM play a crucial role in enhancing transparency and promoting responsible AI usage, they currently exist only as a community norm and are not proactively aligned with privacy and GAI-specific laws and regulations.

\subsection{Broader Impact for Stakeholders}

\subsubsection{GAI app developers}
Previous research has indicated that developers tend to prioritize the development of system-specific functionalities, often neglecting privacy considerations as a primary concern ~\cite{loser2014security,balebako2014improving,li2018coconut}. Our study contributes to the creation of more accurate privacy labels, thereby relieving developers from the time-consuming task of creating these labels manually.
In contrast to the self-declaration approach traditionally required for generating privacy labels, our work shows better usability and compliance.
Furthermore, collaborative development of GAI apps is commonplace, where individual developers may not retain all tool-specific details in memory.
The privacy labels generated by our \texttt{Repo2Label} framework provide developers with a reference for understanding the implementation and rationale behind each privacy practice. 
This facilitates better communication and reduces the overhead for developers who need to collaborate with others.
By automating the generation of privacy labels with \texttt{Repo2Label}, timely updates and adherence to legal regulations are ensured. 
% 之前的研究工作表明，开发人员倾向于开发系统特有的功能，隐私并不被视为他们的主要任务。~\cite{loser2014security,balebako2014improving}。我们的研究有助于创建更精确的隐私标签，从而把开发者从创建隐私标签这项消费时间的工作中解脱出来。对比之前需要开发者手动填写表单来生成隐私标签的形式，我们的工作能够避免开发人员因为对术语的错误理解从而造成生成的隐私标签不准确的问题。同时，GAI工具通常由多个开发者合作开发。在多位开发者协作编码的背景下，单独的开发者并不能记住所有的工具细节。我们\texttt{Repo2Label}框架生成的隐私标签可通过查看"Floating Bubble"了解label答案的参考依据，从而帮助开发人员形成对源代码的整体理解。这使得开发者不必在项目人员之间的沟通消耗时间，而能更专注于开发系统功能。尤其在更新了applications的部分代码从而需要引起了隐私标签的更改时，开发者不需要查看所有的代码细节。借助\texttt{Repo2Label}自动化生成隐私标签，保证了隐私标签的及时性。我们的GAI隐私标签也有助于开发者能开发出更加符合法律法规、更受用户欢迎的工具。

\subsubsection{End-users}
The concise and easily readable format of GAI privacy labels facilitates users in the timely grasping of the privacy practices of GAI apps. 
In comparison to lengthy and academically demanding privacy policies, these privacy labels lower the barrier for ordinary users to understand GAI apps, due to their comprehensibility and acceptability.

While the aim of developing GAI is to benefit a broader demographic, the increasing complexity of privacy practices in these apps can create additional understanding barriers.
Our approach potentially broadens the accessibility of GAI apps to a more diverse range of users. This includes individuals who may not possess advanced literacy skills as well as those who face challenges related to reading disorders. By making privacy practices more transparent and understandable, we can ensure that a broader demographic can benefit from the development of these technologies.

% Additionally, it helps users recognize that there are differences between various  GAI tools, assisting them in making informed decisions about which tools to use.
% 隐私标签以简洁易读的形式有助于用户迅速了解GAI工具的功能，从而在一定程度上改变了用户对GAI总是以黑盒子形式工作的固有印象。由于隐私标签的易理解和可接受性，普通用户对于GAI工具的理解门槛降低了。与冗长且需要大学水平教育理解的隐私政策相比，用户无需具备专业知识即可理解隐私标签，从而使得用户不再感到对AI工具感到困惑。用户可在每次使用AI工具前花费较少时间查看隐私标签，以快速了解该工具的基本信息和披露的隐私权利。借助隐私标签所提供的信息，用户能够进行不同GAI工具之间的比较，并作出相应的使用抉择。

\subsubsection{Regulator}
The rapid advancement of GAI has introduced new risks, positioning regulators as pivotal players in addressing these issues. 
\texttt{Repo2Label} aims to respond to high-level requirements about transparency and privacy disclosure presentation.
For instance, legislation passed in Singapore~\cite{SingaporeGAI} mandates that GAI systems provide model information to downstream users in a format akin to privacy labels, which states \textit{``End-users need greater understanding of content provenance across the content lifecycle and to learn to utilise tools to verify for authenticity.''}. 
Additionally, we advocate for regulators to be more agile in this dynamic market and to provide more proactive UPS solutions to facilitate practitioners in complying with regulations.

% According to Table \ref{table:Tallies}, various regulations have outlined specific demands for watermarking technology. However, our investigation reveals that the majority of GAI apps do not mark their generated content as AI-produced, posing potential copyright and compliance risks. Our design for privacy labels offers a more transparent disclosure of copyright information while also providing basic information about the GAI apps, data rights, and associated risks. Privacy labels facilitate the regulatory activities of regulators for a given GAI app.

\subsection{Broader Impact for the Ecosystem}
% Industry (Hugging Face / GitHub)
Privacy labels can enhance data controllers' internal compliance routines~\cite{novovic2022privacy}. Hence, GAI privacy labels contribute to enhancing the security of the entire open-source software ecosystem. Our privacy label design serves as a reference model for building a responsible AI. Drawing from research in related fields, open-source projects with Model Cards have seen a significant increase in downloads~\cite{liang2024s}, primarily due to improved transparency. This provides a theoretical basis for the expectation that GAI apps with privacy labels in the open-source community will also experience higher user adoption.
Our regulation-driven privacy label ensures that privacy practices and technologies are aligned with current legal standards.

\subsection{Thread to Validity}
1) Internal Validity:
We follow OpenAI's official API to access the aforementioned GPT models in our experiments. We evaluate the performance of our framework on the GAI app dataset. The results from GPT models may deviate due to the probabilistic nature of the model. We have taken extensive measures to mitigate this risk, including providing references when generating labels and implementing a verification step to ensure accuracy. We believe that with the continued development of more advanced foundation models, this issue can be further alleviated.

2) External Validity:
Our privacy label design incorporates common requirements from various regulations to emphasize the current regulatory focus on generative AI tools. However, for our privacy labels to be deployed in individual countries, they must adhere to specific local regulations. In addition, although our framework is evaluated on a dataset based on GitHub code repositories, the core analysis focuses on individual code files, making it can be easily generalized to other code repositories.

% \subsection{Future Work}
% % 我们的GAI隐私标签设计是综合参考了多项法规而提出来的。后续想要在各个国家实际应用，还需要结合本土化的法律法规，增加一些项关键的隐私标签项。
% Add or remove this section to tailor the 13-page limit.

%% file: 8_Conclusion.tex
\section{Conclusion}~\label{sec:7conclusion}
GAI apps have greatly facilitated and enriched our daily lives, yet have raised concerns about transparency in their privacy practices.
Traditional privacy policies often fail to effectively communicate essential privacy information due to their complexity and length, and open-source community developers often neglect privacy practices even more.
Only 12.2\% of examined open-source GAI apps provide a privacy policy.
In this paper, we propose a regulation-driven GAI privacy labels and introduce \texttt{Repo2Label}, a novel framework for automatically generating these labels based on code repositories.
Our user study indicates endorsement of the proposed GAI privacy label design. Additionally, \texttt{Repo2Label} achieves a precision of 0.81, recall of 0.88, and F1-score of 0.84 under optimal settings (GPT-4o and verification) based on the benchmark dataset, significantly outperforming the developer self-declared privacy notices.
Our findings suggest that \texttt{Repo2Label} could serve as a significant tool for bolstering the privacy transparency of GAI apps and make them more practical and responsible.

% We share our code and dataset for researchers to conduct future  privacy label studies.